\documentclass[a4paper,11pt]{article}

\usepackage{jheppub} 
\usepackage{braket}
\usepackage[T1]{fontenc} 
\usepackage{comment}

\subheader{\begin{flushright}
IFT-UAM/CSIC-26-157\\
\end{flushright}}

\title{From bulk flows to boundary generators and costs}

\author[a,b]{Rafael Carrasco,}
\author[c]{Carlos Pérez-Pardavila}

\affiliation[a]{Instituto de Física Teórica UAM/CSIC, Calle Nicolás Cabrera 13-15, Madrid 28049, Spain}
\affiliation[b]{Theory Group, Weinberg Institute, Department of Physics, University of Texas at Austin,\\
2515 Speedway, Austin, Texas 78712, USA.}
\affiliation[c]{Department of Physics and Beyond: Center for Fundamental Concepts in Science, Arizona State University, Tempe, AZ 85281, USA.}

\emailAdd{rafael.carrasco@ift.csic.es}
\emailAdd{cperezpa@asu.edu}

\abstract{Holographic complexity admits a broad family of bulk observables, yet their relation to boundary state-preparation costs remains unclear. We propose a local prescription that associates suitable bulk observables with quadratic cost geometries on families of states prepared by Euclidean sources. Building on the holographic 
transport of bulk Hamiltonian flows to Euclidean preparation sources, we use the Hessian of the transported generator at a stationary reference state to define a candidate cost metric. When this Hessian is positive definite, the minimised preparation energy reproduces the change in the bulk observable to quadratic order. We characterise the resulting preparation penalties relative to the Fubini-Study metric and compute them for the maximal volume in scalar perturbations of AdS$_3$. At large-conformal dimension, we recover the known proportionality between cost and Fubini-Study metrics. At finite dimension, we obtain positive but generally unequal penalties for source variations that displace the bulk field and its conjugate momentum, breaking this proportionality even for a single scalar operator. We further study a deformation of volume by quadratic scalar and gradient terms in the large-dimension regime, determining how the couplings modify the penalties and constrain positivity. These additions selectively change field-displacement costs, with the gradient term introducing mode dependence. Our results provide an explicit local connection between the choice of bulk observables and the assignment of boundary preparation penalties.}

\begin{document}

\maketitle

\section{Introduction}

A central challenge in holography is to relate bulk geometric observables to quantum properties of boundary states. The AdS/CFT correspondence provides a framework for addressing this question by relating gravitational theories in asymptotically anti-de Sitter spacetime to conformal field theories on their boundaries. A landmark example is the Ryu-Takayanagi formula \cite{Ryu:2006bv,Ryu:2006ef} and its covariant extension \cite{Hubeny:2007xt}, which identifies the entanglement entropy of a boundary subregion with the area of an appropriate bulk extremal surface. Beyond providing a prescription for computing entropy, this relation gives a microscopic interpretation to a geometric quantity. It also suggests a broader strategy: understanding information-theoretic properties of boundary quantum states through bulk geometry.

The first law of entanglement provides a particularly instructive example of how this geometric identification can reveal bulk dynamics. For an infinitesimal perturbation of a reference state, it relates the variation of entanglement entropy to the variation of the expectation value of the modular Hamiltonian, $\delta S_A = \delta \braket{ K_A}$, where $K_A = - \log \rho_A$ is held fixed at the reference reduced state. In holographic theories with Einstein gravity duals, requiring this relation for all ball-shaped boundary regions and their Lorentz transforms, for perturbations around the CFT vacuum, implies the linearised Einstein equations for AdS \cite{Lashkari:2013koa,Faulkner:2013ica}. Thus, entanglement not only provides a geometric probe but also constrains the gravitational dynamics. This example suggests a useful route to extending the holographic dictionary, namely, characterising bulk observables through their response to state deformations and identifying the boundary generators associated with those variations.

The prolonged growth of black-hole interiors motivates extending this dictionary beyond entanglement entropy. For a two-sided AdS black hole dual to a thermofield double state, the volume of maximal slices through the Einstein-Rosen bridge continues to grow under boundary time evolution, whereas the entanglement entropy between the two boundary theories remains unchanged. This disparity motivated proposals that it is quantum computational complexity which may provide boundary descriptions of the growing interior \cite{Susskind:2014rva,Susskind:2014moa, Stanford:2014jda}. State complexity measures the minimum number of elementary gates, or unitary operations, needed to prepare a target state from a chosen reference state, with a prescribed accuracy. More importantly, complexity is expected to continue to increase long after other observables have equilibrated, making it a promising candidate for capturing the prolonged evolution of the black hole interior.

The quantum mechanical definition of complexity is a precise one and depends on both a reference and target state. In a quantum field theory, things are more subtle. Nielsen's geometric approach recasts this circuit optimisation as a continuous variational problem on the space of unitary transformations \cite{Nielsen:2006cea,Dowling:2006tnk}. A circuit is represented by a path in the space of states, whose tangent is induced by the instantaneous generator, while a \textit{cost function} assigns a weight to each such direction. Minimising the accumulated cost gives the complexity of a transformation. A related way, closer to the Euclidean preparation of states used below, is a path integral optimisation, where the cost of a Euclidean path integral is taken to be a functional of its background metric, the Liouville action in 2d and its minimisation yields a hyperbolic metric reminiscent of a time slice of AdS \cite{Caputa:2017urj,Caputa:2017yrh}. This procedure can itself be formulated as a circuit complexity problem \cite{Camargo:2019isp}. Different choices of generators and penalties therefore define different preparation problems and associated geometries. This freedom is central to the holographic interpretation: identifying a bulk observable with complexity requires specifying which boundary notion of preparation cost it represents.

On the gravitational side, there is a similar freedom, reflected in the multiple proposals for the duals of complexity that have appeared in recent years. For example, the complexity=volume proposal associates boundary complexity with the volume of a maximal spacelike bulk hypersurface anchored on a boundary time slice \cite{Stanford:2014jda,Susskind:2014rva}. Similarly, the complexity=action proposal associates complexity to the gravitational action evaluated on the corresponding Wheeler-DeWitt patch \cite{Brown:2015bva,Brown:2015lvg}. Later, the complexity=anything framework extends this landscape to a broad family of diffeomorphism-invariant bulk observables \cite{Belin:2021bga,Belin:2022xmt}. In its codimension-one construction, these observables take the form
\begin{equation} \label{eq:complexity-anything}
    O_{F_1,\Sigma_{F_2}}(\Sigma_{CFT}) \sim \int_{\Sigma_{F_2}} d^d\sigma \,\sqrt{h} F_1(g_{\mu\nu};X^\mu),
\end{equation}
where the spacelike bulk hypersurface $\Sigma_{F_2}$ is selected by extremising a second functional
\begin{equation} \label{eq:complexity-anything-surface}
    \delta_X \int_{\Sigma} d^d\sigma \, \sqrt{h} \,F_2(g_{\mu\nu};X^\mu).
\end{equation}
Here, $\Sigma_{\rm CFT}$ is the boundary time slice, $X^\mu(\Sigma)$ describes the embedding of the bulk hypersurface, and $h$ is the determinant of the induced metric. The variation is taken with the boundary anchoring held fixed. The functions $F_1$ and $F_2$ are independently chosen scalars constructed from the bulk geometry and the embedding. For example, the choice of $F_1 = F_2=1$ recovers the complexity=volume proposal. For suitable choices of $F_1$ and $F_2$, this family of observables exhibits the late time linear growth and the switchback effect in black hole geometries \cite{Belin:2021bga,Belin:2022xmt,Myers:2024vve,Jorstad:2023kmq,Caceres:2025myu}. These behaviours provide useful consistency checks, but do not by themselves specify the boundary cost function associated with each observable. Although the freedom to choose a bulk functional is suggestive of the freedom to assign a boundary cost function, a systematic correspondence between these two remains to be established.

In order to make this correspondence precise, we need to answer two questions. The first one is how a bulk observable can be represented as a boundary functional. The second one is whether that functional can be realised as the optimal cost of preparing a state from a fixed reference. We observe that providing an answer to the first question does not automatically answer the second one: identifying an observable and its variation does not necessarily assign costs to all admissible directions or establish minimising preparation paths. The holographic transport of Hamiltonian flows provides an approach to the first question \cite{Belin:2022xmt,Belin:2018fxe,Belin:2018bpg}. Cost functions have previously been associated with specific bulk quantities, such as the volume through the York flow \cite{Belin:2018bpg}, the gravitational action through the first law of complexity \cite{Bernamonti:2019zyy,Bernamonti:2020bcf}, and functionals on bulk subregions in the cost of holographic path integrals \cite{Chandra:2022pgl}. Conversely, for circuits of conformal transformations in two-dimensional CFTs, the Fubini-Study cost has been expressed through bulk geometric quantities \cite{Erdmenger:2021wzc,Erdmenger:2022lov} and penalty factors have been introduced directly in the boundary complexity geometry \cite{Baiguera:2025uss}. Here we use the symplectic transport as the starting point for addressing the second question as well. Our focus is a local prescription that associates suitable boundary generators with preparation cost geometries, together with explicit calculations of the penalties selected by different bulk observables.

The starting point is the holographic relation between the bulk and boundary phase-space symplectic structures. On the boundary side, Euclidean path integrals with sources prepare a family of states whose overlaps determine a K\"ahler structure. In the semiclassical regime, the symplectic form arising from this structure agrees with the covariant phase-space symplectic form of the dual bulk theory for corresponding on-shell deformations \cite{Belin:2018fxe}. This identification allows us to relate bulk flows to boundary flows. If in addition one relates a bulk observable to its Hamiltonian flow, one can identify a corresponding boundary flow. A useful precedent is provided by the maximal-volume observable, whose Hamiltonian vector field is the York deformation. Furthermore, its symplectic pairing with an allowed perturbation gives the variation of the volume observable \cite{Belin:2018bpg}. Together with an appropriate cost prescription, this relation underlies a first law of holographic complexity, with extensions to higher derivative gravity and semiclassical dilaton gravity \cite{Bernamonti:2019zyy,Bernamonti:2020bcf,Pedraza:2021mkh,Pedraza:2021fgp,Carrasco:2023fcj}. Belin et al. extended this strategy to general diffeomorphism-invariant bulk observables in the complexity=anything framework \cite{Belin:2022xmt}. Their construction identifies the bulk conjugate variation through canonical methods or the Peierls prescription and represents it as a deformation of the Euclidean preparation sources.

We review this construction in a form suited to the subsequent cost analysis. Starting from a differentiable, diffeomorphism invariant bulk observable $W_{\text{bulk}}$, regarded as a ``Hamiltonian'' on the reduced gravitational phase space, its Hamiltonian vector field can be constructed using the Peierls prescription, by taking the difference between the retarded and advanced responses to an infinitesimal deformation of the action \cite{Belin:2022xmt,Peierls:1952cb,Harlow:2019yfa}. The holographic map described above transports this vector field to the boundary, where its action is encoded in variations of sources. Pairing the transported flow with the boundary symplectic form produces a one-form which, under the local symplectic identification, determines the corresponding boundary generator. Schematically, the logic is
\begin{equation}
    W_{\rm bulk}\rightarrow \delta_W^{\rm bulk} \rightarrow \delta_W^\partial\rightarrow H_W^{\rm CFT}.
\end{equation}
Working locally where the transported flow is tangent to the chosen family of states, this construction determines the boundary generator $H_W^{\rm CFT}$, up to an additive constant. It therefore provides a way to identify the boundary counterpart of a bulk observable through its action on holographic data, without committing to a particular preparation cost geometry. It is therefore the input to our proposed cost construction.

We also test the transport construction against the established relation between extremal area and boundary entanglement. In Appendix~\ref{app:B}, we consider a half-line in the vacuum of a two-dimensional CFT and determine the boundary deformation induced by the bulk area flow. Its pairing with the boundary symplectic form reproduces the variation of the vacuum modular Hamiltonian and, through the Ryu-Takayanagi relation, the first law of entanglement. 

We propose a local prescription that uses this boundary generator to define the geometry of preparation costs. We work near a reference state at which the generator is stationary and has positive-definite Hessian. Half of this Hessian supplies a candidate quadratic cost metric on the space of preparation sources. In the quadratic approximation, the preparation problem reduces to minimising an energy functional over paths joining the reference and target states. For a fixed parameter interval and an appropriate normalisation, the minimum reproduces the change in the boundary generator and hence in the bulk observable to quadratic order. This agreement holds by construction. The physical content of the prescription lies in the preparation penalties selected by observables, which we characterise relative to the Fubini-Study metric: directions of equal Fubini-Study norm have equal leading changes in fidelity, allowing their preparation costs to be compared. The linearised transported flow also determines a compatible complex structure, which relates the optimal preparation directions to that flow through a rotation and a frequency-dependent rescaling. 

We first apply the cost prescription to maximal volume in the presence of a perturbative scalar source. The large-conformal dimension scalar example was previously analysed in \cite{Belin:2018bpg}. We revisit it as a consistency check, recovering a cost metric proportional to the Fubini-Study metric, so that all directions within the scalar sector have the same relative penalty. We then extend the calculation to finite conformal dimension in the $x$-independent scalar sector around the AdS$_3$ vacuum and compute the resulting principal penalties. These remain positive but are generally unequal for variations of the real and imaginary parts of the preparation source, which displace the bulk scalar field and its conjugate momentum, respectively. The two preparation directions therefore carry different costs even when they produce the same leading change in fidelity with the reference state. Thus, a single scalar operator already produces a cost geometry that is not proportional to the Fubini-Study metric, with the common large-dimension penalty recovered at leading order. This extension makes explicit how the penalties change beyond the earlier approximation, establishes their positivity and identifies their meaning in terms of boundary preparation sources.

To examine how the choice of bulk observables changes these penalties, we next deform volume by adding a quadratic scalar field and spatial gradient terms. In the large-conformal dimension regime, we compute the corresponding boundary generator, cost metric and principal penalties. We also determine the restrictions on the couplings required for positivity. These additions modify the field displacement penalties while leaving momentum-displacement penalties unchanged at the order considered. The gradient term also introduces mode dependence, allowing different source profiles to carry different relative costs. Together, the two examples give a concrete realisation of the proposed relation between bulk observables and boundary preparation penalties. Different admissible bulk functionals can select different local preparation geometries, providing a way to organise the freedom present in holographic complexity. The construction fixes these costs near the reference state; extending the correspondence beyond the quadratic approximation requires additional input.

The remainder of this paper is organised as follows. Section \ref{sec2:holographic_transport} reviews Euclidean-source state preparation, the holographic identification of the bulk and boundary symplectic forms, the Peierls construction and the transport prescription of~\cite{Belin:2018fxe,Belin:2018bpg,Belin:2022xmt}. In Section~\ref{sec3: CV} we apply this prescription to the observable of volume of a maximal slice in the bulk, with and without scalar field, which provides the starting point for the cost analysis. Section~\ref{sec4: flow_to_cost} presents our local cost construction and characterises its penalties with respect to the Fubini-Study metric. Here, we also recover the volume result in the presence of a large-conformal dimension scalar field and compute the finite dimension penalties and study the scalar deformation of the volume. Section~\ref{sec:discussion} discusses the implications, limitations and future directions of this construction. Appendix~\ref{app1:Review} reviews the covariant phase space formalism. Appendix~\ref{app:B} applies the transport prescription to the extremal area and recovers the first law of entanglement entropy. Finally, Appendix~\ref{app:FG_York} restores Fefferman-Graham gauge for the York deformation.

\section{Holographic transport of bulk observables}\label{sec2:holographic_transport}

In this section, we review the holographic transport of bulk observables, following the relation between bulk Hamiltonian flows and boundary source deformations described in \cite{Belin:2018fxe, Belin:2022xmt}. We organise these ingredients in a form convenient for the subsequent construction of cost functions and observable-dependent penalties. Let $W_{\rm bulk}$ be a differentiable, gauge-invariant observable on the bulk phase space. Its associated Hamiltonian vector field $\delta_W^{\rm bulk}$ defined as in \eqref{eq: hamilton_eq} of Appendix~\ref{app1:Review} satisfies
\begin{equation}
	\delta W_{\rm bulk}=\Omega_{\rm bulk}(\,\cdot\,, \delta_W^{\rm bulk}).
	\label{eq:bulk_hamiltonian_vector_intro}
\end{equation}
The equality of the bulk and boundary symplectic forms allows this flow to be represented on a suitable family of Euclidean source states and its generator to be expressed in terms of boundary preparation data.

The review proceeds in four steps. Section~\ref{subsec:sources} describes how Euclidean path integrals with sources prepare a family of CFT states that at large $N$ are dual to classical bulk solutions. Section~\ref{subsec:fs_boundary_symplectic} shows that this family of states carries a K\"ahler structure whose symplectic form $\Omega_\partial$ equals the bulk symplectic form \cite{Belin:2018fxe}. In Section~\ref{subsec:peierls} we review how the Peierls prescription constructs $\delta_W^{\rm bulk}$. Finally, in Section~\ref{subsec:transport} we assemble these ingredients: whenever the bulk flow admits a representation by allowed preparation-source variations $\delta_W^{\partial}$, its boundary generator satisfies $\delta H_{W}^{\rm CFT}=\Omega_\partial(\,\cdot\,, \delta_W^{\partial})$. Here $H_{W}^{\rm CFT}$ denotes a functional on the chosen family of source prepared states. This formulation supplies the boundary flows used in subsequent cost constructions.

\subsection{Euclidean-source states and holographic data}\label{subsec:sources}

We begin by reviewing the class of CFT states on which the boundary phase-space description will be based \cite{Botta-Cantcheff:2015sav,Skenderis:2008dg,Marolf:2017kvq}. Let $\mathcal M_E$ denote a $(d+1)$-dimensional asymptotically Euclidean AdS bulk spacetime, with conformal boundary
\begin{equation}
	\mathcal{B}_E = \partial \mathcal M_E
\end{equation}
on which the CFT is defined. We assume that the preparation manifold $\mathcal B_E$ admits a reflection-symmetric Euclidean time coordinate $\tau$ and a separating codimension-one slice $\Sigma_{\mathrm{CFT}}$ at $\tau = 0$. Cutting along this slice divides $\mathcal{B}_E$ into the lower and upper preparation regions,
\begin{equation}
	\mathcal B_-=\{\tau<0\},
	\qquad
	\mathcal B_+=\{\tau>0\}.
\end{equation}

The basic holographic input is the Euclidean AdS/CFT relation between the CFT partition function with sources and the bulk partition function. For a collection of CFT operators $\mathcal O_a$ sourced by $\lambda_a$, we write
\begin{equation}
	Z_{\rm CFT}[\lambda]
	=  Z_{\rm CFT}[0]\left\langle \exp\left[ -\int_{\mathcal B_E} d^dx\,\sqrt{\gamma}\, \lambda_a(x)\mathcal O_a(x)\right]
	\right\rangle =\exp\left[-I_E^{\rm grav}[\lambda]\right].
	\label{eq:euclidean_ads_cft_dictionary}
\end{equation}
Here $\gamma_{ij}$ is the Euclidean boundary metric, the index $a$ labels the sourced operators, and the brackets denote a normalised Euclidean expectation value at $\lambda=0$. The metric dependence is displayed separately, with $\gamma_{ij}$ serving as the source for the stress tensor. The last relation holds at leading semiclassical order when a single bulk saddle dominates, and $I_E^{\rm grav}$ denotes the renormalised bulk action evaluated on that saddle.

With the source convention in \eqref{eq:euclidean_ads_cft_dictionary}, varying a scalar source at fixed boundary metric yields the one-point function for  the associated operator
\begin{equation}
	\langle \mathcal O_a(x)\rangle_{\lambda}=\frac{1}{\sqrt{\gamma(x)}} \frac{\delta \mathcal W_{\rm CFT}[\lambda]} {\delta \lambda_a(x)} \quad \text{with}\quad \mathcal W_{\rm CFT}[\lambda]
	\equiv -\log Z_{\rm CFT}[\lambda]
	\simeq I_E^{\rm grav}[\lambda].
	\label{eq:one_point_from_source_variation}
\end{equation}
For the boundary metric, we use the standard convention \cite{deHaro:2000vlm}, where the previous expression is multiplied by a factor of $-2$.

Thus, the relevant boundary data are naturally organised into source/one-point function pairs $(\lambda_a(x),\langle \mathcal O_a(x)\rangle_{\lambda})$.

In the bulk, these pairs are encoded in the near-boundary expansion of the dual fields \cite{deHaro:2000vlm,Skenderis:2002wp}. In Fefferman--Graham gauge,
\begin{equation}
	\begin{split}
		ds^2&=\frac{\ell^2}{z^2}\bigl[dz^2+g_{ij}(z,x)\,dx^idx^j\bigr],\quad g_{ij}(z,x)=\gamma_{ij}(x)+\dots+z^d g_{(d)ij}(x)+\dots,\\
		\phi_a(z,x)&=z^{d-\Delta_a}\bigl(\lambda_a(x)+\dots\bigr)+z^{\Delta_a}\bigl(A_a(x)+\dots\bigr),
	\end{split}
	\label{eq:fg_expansion}
\end{equation}
with the conformal boundary at $z=0$ and $\Delta_a$ the conformal dimension of the scalar field. The leading coefficients $\gamma_{ij}$ and $\lambda_a$ are the sources of $T^{ij}$ and $\mathcal O_a$, while $g_{(d)ij}$ and $A_a$ determine $\braket{T_{ij}}$ and $\braket{\mathcal O_a}$, up to local terms fixed by holographic renormalisation. Logarithmic terms appear in the metric for even $d$, and in the scalar when $\Delta_a-d/2$ is an integer.

Conversely, Euclidean sources can be used to prepare states in the CFT \cite{Botta-Cantcheff:2015sav,Skenderis:2008dg,Marolf:2017kvq}. A ket is obtained by performing the Euclidean path integral over the lower half $\mathcal B_-$ with sources turned on:
\begin{equation}
	|\lambda^-\rangle = \mathcal T_E
	\exp\left[ -\int_{\mathcal B_-} d^d x\,\sqrt{\gamma}\,  \lambda^-_{a}(x)\mathcal O_a(x) \right] |0\rangle .
	\label{eq:source_prepared_ket}
\end{equation}
Here $|0\rangle$ denotes the Hartle-Hawking vacuum, i.e. the state prepared by the same path integral with all non-trivial sources set to zero, dual to the Hartle-Hawking state of the bulk, and $\mathcal T_E$ is Euclidean time ordering. Similarly, a bra is prepared by a path integral over the upper half $\mathcal B_+$,
\begin{equation}
	\langle \lambda^+|=\langle 0|\mathcal T_E \exp\left[-\int_{\mathcal B_+} d^dx\,\sqrt{\gamma}\, \lambda^+_{a}(x)\mathcal O_a^{\dagger}(x)
	\right].
	\label{eq:source_prepared_bra}
\end{equation}
The bra is the Hermitian conjugate of the ket when the sources obey the reality condition
\begin{equation}\label{eq:reality_condtion}
	\lambda^-(-\tau,x)=[\lambda^+(\tau,x)]^*.
\end{equation}

We denote by $\mathcal S$ the manifold of preparation sources. A point in $\mathcal{S}$ is a collection of complex source profiles $\lambda^-=\{\lambda^-_a\}$ on $\mathcal{B}_-$, including the boundary metric when it is varied. Likewise, a tangent vector is a collection of variations $\delta \lambda^-=\{\delta\lambda_a^-\}$. Since the Euclidean sources are naturally complex, we work on a complexified source manifold. Via \eqref{eq:reality_condtion}, the sources $\lambda^-$ and $\lambda^+$ can then be identified with holomorphic and antiholomorphic coordinates, respectively.

In the large-$N$ semiclassical regime, a source-prepared state is dominated by a classical Euclidean saddle. Cutting it along a bulk slice anchored at $\tau=0$ and continuing analytically gives initial data for a Lorentzian solution $\Phi_\lambda$, so that, at least locally, each source configuration $\lambda$ determines a state $\ket{\lambda}$ and a bulk solution $\Phi_\lambda$ \cite{Skenderis:2008dg,Skenderis:2008dh,Belin:2020zjb}, whose near-boundary expansion~\eqref{eq:fg_expansion} encodes the pairs $(\lambda_a,\braket{\mathcal O_a}_\lambda)$.

\subsection{Fubini-Study geometry and the boundary symplectic form}
\label{subsec:fs_boundary_symplectic}

We now review the Fubini-Study geometry inherited by the family of Euclidean-source states introduced above from projective Hilbert space. In the semiclassical regime, the associated K\"ahler form provides the boundary symplectic structure. Its equality with the renormalised covariant phase-space symplectic form of the bulk theory, evaluated on corresponding bulk and boundary variations, was established in \cite{Belin:2018fxe}. We express this identification in source/one-point function variables for use in the subsequent transport and cost constructions.

The source-prepared states $|\lambda^-\rangle$ constructed above define a family of rays in projective Hilbert space. At leading order in the $1/N$ expansion, and for single trace operators, they are coherent states of the dual bulk fields, which are labelled by points of the classical phase space \cite{Botta-Cantcheff:2015sav,Marolf:2017kvq}. This is what underlies the identification with the bulk symplectic structure. The Kähler property used below relies only on the holomorphic dependence of the states on the sources. By construction, the ket \eqref{eq:source_prepared_ket} depends only on $\lambda^-_a$ and the bra \eqref{eq:source_prepared_bra} only on $\lambda^+_a$
\begin{equation}
	\partial_{\lambda^+}\ket{\lambda^-}=0,\qquad\partial_{\lambda^-}\bra{\lambda^+}=0.
\end{equation}
The projective space comes with a natural metric induced by the inner product, the Fubini-Study metric. For a family of unnormalised states $|\psi\rangle$,
it can be written as
\begin{equation}
	ds^2_{\rm FS} = \frac{\langle \delta\psi|\delta\psi\rangle}{\langle\psi|\psi\rangle}-  \frac{\langle \delta\psi|\psi\rangle
		\langle\psi|\delta\psi\rangle}
	{\langle\psi|\psi\rangle^2}.
	\label{eq:fs_metric_general}
\end{equation}
When the states depend holomorphically on complex coordinates, this is a K\"ahler metric with K\"ahler potential $\log\braket{\psi|\psi}$. For source prepared states, it reads
\begin{equation}
	\mathcal K[\lambda^+,\lambda^-] =    \log \langle \lambda^+|\lambda^-\rangle
	= \log Z_{\rm CFT}[\lambda^+,\lambda^-].
	\label{eq:kahler_potential_sources}
\end{equation}
Here, $Z_{\rm CFT}[\lambda^+,\lambda^-]$ is the CFT path integral over $\mathcal B_E$ with sources $\lambda^-$ inserted on $\mathcal{B}_-$ and $\lambda^+$ on $\mathcal{B}_+$.

Associated with the K\"ahler metric is the K\"ahler form
\begin{equation}
	\begin{split}
		\Omega_{\rm FS}(\delta_1,\delta_2) =& -i\partial\bar\partial\mathcal K=i\int_{\mathcal B_-} dx_-\sqrt{\gamma_-}\int_{\mathcal B_+} dx_+\sqrt{\gamma_+}G_{a\bar{b}}(x_-,x_+)\\
		&\times \left(\delta_1\lambda^+_b(x_+^T)\delta_2\lambda^-_a(x_-)-\delta_2\lambda^+_b(x_+^T)\delta_1\lambda^-_a(x_-)\right).
		\label{eq:fs_kahler_form}
	\end{split}
\end{equation}
Here $\partial$ and $\bar\partial$ act on the holomorphic and antiholomorphic source coordinates, respectively, $x^T=(-\tau,\vec{x})$ and $G_{a\bar b}(x_-,x_+)$ is the Euclidean two-point function in the source-deformed state. With the metric normalisation in \eqref{eq:fs_metric_general}, this convention gives $g_{\rm FS}(u,v)=\tfrac12\Omega_{\rm FS}(u,J_{\rm FS}v)$, where $J_{\rm FS}$ is the source-space complex structure, which acts on the sources as
\begin{equation}
	J_{\rm FS}\lambda^-=-i\lambda^-,\qquad J_{\rm FS}\lambda^+=i\lambda^+
\end{equation}

It is useful to rewrite the same object in source/one-point function variables. We define the momentum conjugate to the sources as \cite{Skenderis:2002wp}
\begin{equation}
	\Pi_a(x)\equiv -\frac{\delta \mathcal K}{\delta\lambda^+_a(x)}, \qquad x \in \mathcal{B}_+
\end{equation}
By \eqref{eq:one_point_from_source_variation}, this gives
\begin{equation}\label{eq:pi_vevs}
	\begin{split}
		\Pi_a= \sqrt{\gamma}\braket{\mathcal O_a}_\lambda\quad &\text{for scalar sources},\\
		\Pi^{ij}=-\frac{1}{2}\sqrt{\gamma}\braket{T^{ij}}_\lambda\quad &\text{for metric sources}.
	\end{split}
\end{equation}

Using \eqref{eq:pi_vevs} and that the variation of $\braket{\mathcal{O}_a}$ is $\delta \braket{\mathcal{O}_a}=-\int d^dx_-\sqrt{\gamma}G(x_-,x_+)\delta\lambda_a^-(x_-)$, the K\"ahler form can be written in canonical source/one-point function form as
\begin{equation}
	\Omega_{\partial}(\delta_1,\delta_2)
	\equiv
	\Omega_{\rm FS}(\delta_1,\delta_2)
	=
	i
	\int_{\mathcal B_+}
	d^d x\,
	\left(\delta_1^-\Pi\delta\lambda^+_2 -\delta_2^-\Pi\delta \lambda^+_1\right),
	\label{eq:boundary_symplectic_canonical_form}
\end{equation}
where $\delta^\pm$ denotes the part of a variation $\delta$ that acts only on the sources on $\mathcal B_\pm$, so that $\delta=\delta^++\delta^-$. In this expression, for notational simplicity we have omitted the index $a$ labelling the sources and conjugate momenta. Nevertheless, the sum over $a$ is understood. Equivalently, suppressing the density notation,
\begin{equation}\label{eq: boundary_symplectic_form}
	\Omega_{\partial}(\delta_1,\delta_2)
	=
	i
	\int_{\mathcal B_+}
	d^d x\,\sqrt{\gamma}\,\left(\delta_1^-\braket{\mathcal{O}}\delta\lambda^+_2 -\delta_2^-\braket{\mathcal{O}}\delta\lambda^+_1 \right),
\end{equation}
with the understanding that variations of the measure and the appropriate tensor contractions are included when the metric is among the sources. Two features of both \eqref{eq:boundary_symplectic_canonical_form} and \eqref{eq: boundary_symplectic_form} will be important later. First, only the response of $\Pi_a$ and $\braket{\mathcal O_a}$ to the lower-half sources contributes \textit{i.e.} $\delta\Pi_a\rightarrow\delta^-\Pi_a$. Second, on the real section~\eqref{eq:reality_condtion}, $\mathcal K=\log\braket{\lambda|\lambda}$ is real, so $\Omega_\partial=-i\partial\bar\partial\mathcal K$ is real.

The key result of \cite{Belin:2018fxe} (see also \cite{Czech:2023zmq}) is that the boundary K\"ahler form and the bulk symplectic form agree on variations related by the holographic dictionary:
\begin{equation}
	\Omega_{\rm bulk}(\delta_1\Phi,\delta_2\Phi)
	=
	\Omega_{\partial}(\delta_1\lambda,\delta_2\lambda).
	\label{eq:bulk_boundary_symplectic_equality}
\end{equation}
Here  $\delta_i\lambda$ are variations of the preparation
sources, and $\delta_i\Phi$ are the corresponding bulk
perturbations satisfying the linearised equations of motion. Thus, $\Omega_{\rm bulk}$ is defined on the bulk phase space, while $\Omega_\partial$ is defined on the source manifold; the equality relates their evaluations on the corresponding variations within the chosen semiclassical family.

Let us sketch the mechanism behind \eqref{eq:bulk_boundary_symplectic_equality} referring to \cite{Belin:2018fxe} for further details. By \eqref{eq:kahler_potential_sources} and \eqref{eq:euclidean_ads_cft_dictionary}, $\mathcal K=-I_E^{\rm grav}$. On-shell, $\delta L=d\theta$ (as reviewed in Appendix \ref{app1:Review}) gives $\delta I_E^{\rm grav}=\int_{\mathcal{B}_E}\theta_E$. Since $\delta^\pm$ commute and $\theta_E(\delta^\pm)$ vanishes on $\mathcal{B}_\mp$, one finds $\delta_1^-\delta_2^+I_E^{\rm grav}=\int_{\mathcal B_+}\delta_1^-\theta_E(\delta_2^+)$, while the terms with two variations of the same half cancel in the antisymmetric combination. Hence
\begin{equation}
	\Omega_\partial(\delta_1,\delta_2)=i\left(\delta_1^-\delta_2^+-\delta_2^-\delta_1^+\right) I_E^{\rm grav}=i \int_{\mathcal{B}_+}\omega_E(\delta_1,\delta_2)
\end{equation}
where $\omega_E(\delta_1,\delta_2)=\delta_1 \theta(\delta_2)-\delta_2 \theta(\delta_1)$ is the Euclidean symplectic current. The right-hand side is an integral over $\mathcal{B}_+$ but not on a Cauchy slice, a condition required for this term to be a bulk symplectic form. However, due to the fact that $d\omega_E=0$, applying Stokes' theorem on the half $\mathcal{M}_+$ of the saddle, with $\mathcal{M}_+$ the region enclosed by the $\tau=0$ slice and $\mathcal{B}_+$, then moves the integral to the bulk slice $\Sigma$ at $\tau=0$ up to corner contributions that are assumed to vanish.

For sources obeying \eqref{eq:reality_condtion}, and variations preserving it, the data on $\Sigma$ can be analytically continued to initial data of a real Lorentzian solution. With $t=-i\tau$, the momenta satisfy $\pi_L=i\pi_E$, so that $i\int_\Sigma \omega_E=\Omega_{\rm bulk}$. This establishes \eqref{eq:bulk_boundary_symplectic_equality}, where the factor $i$ appearing in $\Omega_\partial$ is precisely produced by the Wick rotation of the momenta.

The equality \eqref{eq:bulk_boundary_symplectic_equality} underlies the transport of bulk observables to the boundary. Having identified the boundary symplectic form, we now review how to construct the Hamiltonian vector field associated with a given bulk observable using the Peierls prescription.

\subsection{Peierls construction and bulk Hamiltonian vector fields}\label{subsec:peierls}

We now return to the bulk side and construct the phase-space vector field $\delta_W^{\rm bulk}$ associated with the observable $W_{\rm bulk}$.

Equation \eqref{eq:bulk_hamiltonian_vector_intro} is an implicit definition of $\delta_W^{\rm bulk}$. In order to push the flow to the boundary, we need the corresponding perturbation of the bulk fields $\delta_W\Phi$. 

The Peierls construction gives this perturbation by studying the retarded and advanced responses of the bulk fields to a deformation of the action by $W_{\rm bulk}$. More precisely, we choose an off-shell extension $W[\Phi]$ of the phase-space observable $W_{\rm bulk}$, and consider
\begin{equation}
	I_\epsilon[\Phi] =I[\Phi]-\epsilon W[\Phi],
\end{equation}
with $\epsilon$  infinitesimal. Starting from a solution $\Phi_0$ of the undeformed theory, we then construct two solutions of the deformed equations,
\begin{equation}
	\Phi_R= \Phi_0+\epsilon \delta_R\Phi+O(\epsilon^2),
	\qquad \Phi_A =\Phi_0+\epsilon \delta_A\Phi+O(\epsilon^2).
\end{equation}
The subscripts $R$ and $A$ denote retarded and advanced boundary conditions: $\delta_R\Phi$ vanishes on a Cauchy surface $\Sigma_-$ to the past of the support of the deformation, while $\delta_A\Phi$ vanishes on a Cauchy surface $\Sigma_+$ to its future. The Peierls variation is then defined as
\begin{equation}
	\delta_W\Phi = \delta_R\Phi-\delta_A\Phi.
	\label{eq:peierls_variation_summary}
\end{equation}
Since the retarded and advanced perturbations solve the same sourced linearised equations, their difference solves the homogeneous linearised equations and therefore defines a tangent vector to the original phase space. The Peierls theorem \cite{Peierls:1952cb} states that this tangent vector is precisely the Hamiltonian vector field generated by $W_{\rm bulk}$,
\begin{equation}\label{eq:peierls_summary_identity}
	\delta W_{\rm bulk} = \Omega_{\rm bulk}(\,\cdot\,,\delta_W).
\end{equation}
Thus, while the equation \eqref{eq:bulk_hamiltonian_vector_intro} defines the Hamiltonian vector field abstractly, the Peierls construction gives the explicit field variation  $\delta_W\Phi$ that can be expanded near the boundary to read off the induced variations of the CFT sources and one-point functions.

\subsection{The transport prescription}\label{subsec:transport}

We now assemble the ingredients reviewed above into the transport prescription relating bulk observables to boundary source deformations, as described in \cite{Belin:2022xmt}. We organise this prescription in terms of the induced source and response variations, which will be used in the subsequent cost construction. Additionally, we integrate the resulting exact one-form on the chosen local family of source-prepared states. This reconstructs $H_W^{\mathrm{CFT}}$, up to an additive constant, as a functional of boundary preparation data. The examples of Section~\ref{sec3: CV} and Appendix~\ref{app:B} illustrate how this procedure produces explicit boundary expressions for the transported observables.

Let $W_{\rm bulk}$ be a differentiable, gauge-invariant observable on the bulk phase space. The first step is to construct its Hamiltonian vector field $\delta_W^{\rm bulk}$. While this vector field is defined implicitly through its pairing with $\Omega_{\rm bulk}$, the Peierls prescription described above provides a way to compute
the corresponding bulk perturbation. Henceforth, whenever there is no confusion, we will omit the superscripts ``bulk'' and ``$\partial$'' in the variations.

The next step is to determine the boundary deformation corresponding to $\delta_W^{\rm bulk}$. We work locally where this bulk deformation can be represented by allowed variations of the Euclidean preparation sources, and denote the corresponding boundary flow by $\delta_W^\partial$. For any allowed boundary variation $\delta_1^\partial$ and the bulk perturbation $\delta_1$ it induces, the equality \eqref{eq:bulk_boundary_symplectic_equality} gives
\begin{equation}
	\Omega_{\rm bulk}\bigl(\delta_1,\delta_W^{\rm bulk}\bigr)  =\Omega_{\partial}\bigl(\delta_1^\partial,\delta_W^\partial\bigr).
	\label{eq:transport_symplectic_pairing}
\end{equation}
Thus, the symplectic pairing defining the bulk Hamiltonian vector field can be evaluated entirely in terms of boundary preparation data.

Operationally, $\delta_W^{\rm bulk}$ solves the homogeneous linearised equations, and it is determined by initial data $(\delta_W\Phi,\delta_W \pi)$ on the Cauchy slice $\Sigma$. Continuing the bulk perturbation to the Euclidean preparation saddle (with $t=-i\tau$) and expanding it near the asymptotic boundary gives a linearised Euclidean solution on both $\mathcal{B}_\pm$. After restoring Fefferman-Graham gauge when necessary, the asymptotic coefficients determine the variations of the preparation sources $\delta_W \lambda^\pm$ and their associated one-point functions $\delta_W\braket{\mathcal O}$. A real Lorentzian perturbation corresponds to a variation that satisfies the reflection condition \eqref{eq:reality_condtion}. 

Combining the equality of symplectic pairings with the defining property of the bulk Hamiltonian vector field \eqref{eq:bulk_hamiltonian_vector_intro} gives
\begin{equation}
	\Omega_{\partial} \bigl(\delta_1^\partial,\delta_W^\partial\bigr)= \Omega_{\rm bulk} \bigl(\delta_1,\delta_W^{\rm bulk}\bigr)= \delta_1 W_{\rm bulk},
\end{equation}
where $\delta_1$ is the bulk perturbation corresponding to the allowed boundary variation $\delta_1^\partial$. The last expression is the variation of $W_{\rm bulk}$ evaluated on the source-prepared family of solutions. Consequently, the boundary pairing is the pull-back of $\delta W_{\rm bulk}$ to the family of source-prepared states and defines a locally exact one-form and determines a boundary generator $H_W^{\rm CFT}$ satisfying
\begin{equation}
	\delta_1^\partial H_W^{\rm CFT} = \Omega_{\partial}
	\bigl(\delta_1^\partial,\delta_W^\partial\bigr) = \delta_1 W_{\rm bulk}.
	\label{eq:boundary_generator_definition}
\end{equation}
The boundary generator is determined locally on the chosen family of states, up to an additive constant. The prescription reconstructs its dependence on boundary preparation data through the induced flow, without requiring its CFT expression to be known in advance. These boundary flows and generators provide the starting point for the subsequent construction of cost functions and observable dependent penalty factors.

\section{Complexity=Volume}\label{sec3: CV}
We now apply the transport prescription to the volume observable of the ``Complexity=Volume'' (CV) proposal. This calculation has already been discussed in detail in the literature, more concretely in \cite{Belin:2018bpg, Belin:2022xmt}. The reason why we revisit this calculation is two-fold: first, it illustrates the application of Peierls construction and, second, it provides the setup in which we test the cost construction of Section~\ref{sec4: flow_to_cost}.

For simplicity, we will restrict to vacuum AdS$_{d+1}$ in general relativity. In a first stage, we will review the application of the Peierls' formalism in order to find which is the variation of the metric $\delta_V g_{\mu\nu}$ such that 
\begin{equation} \label{eq:deltaV}
	\frac{\delta V}{G_N \ell}=\Omega_{\textup{bulk}}(\delta,\delta_V).
\end{equation}
Then, since the new solution is not in Fefferman-Graham gauge, we restore the gauge through a perturbative change of coordinates, which we compute explicitly in Appendix~\ref{app:FG_York}. This gives the variations of the boundary metric and of the CFT stress tensor, and hence the boundary observable dual to the volume.

As a functional, we will consider the volume of the maximal slice $\Sigma$ divided by $G_N \ell$ 
\begin{equation}\label{eq:generalized vol}
	V=\frac{1}{G_N\ell}\int_{\Sigma} d^d x\sqrt{h},
\end{equation}
where $h$ is the determinant of the induced metric on $\Sigma$. 

First, following Peierls' approach, we start finding the retarded and advanced solutions to the equations of motion associated to the deformed action
\begin{equation}
	S=\frac{1}{16\pi G_N}\int d^{d+1}x\sqrt{-g}\left(R+\frac{d(d-1)}{\ell^2}\right)-\epsilon V.
\end{equation}
Notice that for the unperturbed equations of motion, the solution is vacuum AdS$_{d+1}$, which can be written in the Wheeler-DeWitt form as
\begin{equation}
	ds^2=-d\tau^2+\ell^2\cos^2\left(\frac{\tau}{\ell}\right)d\Sigma^2.
\end{equation}
In these coordinates, the surface of maximal volume is located at $\tau=0$. Based upon this expression, we propose the following ansatz for the retarded solution
\begin{equation}\label{eq: retarded ansatz}
	\begin{split}
		d s^2_R=&-d \tau^2+\left[\Theta(-\tau)\ell^2\cos^2\left(\frac{\tau}{\ell}\right)+\Theta(\tau)\ell^2\cos^2\left(\frac{\tau-\epsilon \beta_R}{\ell}\right)\right]d\Sigma^2.
	\end{split}
\end{equation}
Here $\beta_R$ is a parameter needed for the junction conditions to be satisfied. These conditions impose the following constraints \cite{Israel:1966rt}
\begin{equation}\label{eq:retarded_constraints}
	\begin{split}
		h_{ij}(\tau=0^-)&=h_{ij}(\tau=0^+),\\
		[K_{ij}]-h_{ij} [K]&=-\frac{8\pi \epsilon}{\ell}h_{ij}.
	\end{split}
\end{equation}
where $[A]=A(\tau= 0^+)-A(\tau=0^-)$ and $K_{ij}$ is the extrinsic curvature tensor of $\Sigma$ with future directed normal. In our particular case, the first equation is trivially satisfied as long as $\epsilon \beta_R$ is small, while the second reads
\begin{equation}
	\begin{split}
		\beta_R= \frac{8\pi \ell}{d-1}
	\end{split}
\end{equation}
The advanced solution can be obtained in a similar fashion but with a slight modification of the original ansatz in \eqref{eq: retarded ansatz}. Here, we will consider
\begin{equation}\label{eq: advanced_ansatz}
	\begin{split}
		d s^2_A=&-d \tau^2+\left[\Theta(-\tau)\ell^2\cos^2\left(\frac{\tau-\epsilon\,\beta_A}{\ell}\right)+\Theta(\tau)\ell^2\cos^2\left(\frac{\tau}{\ell}\right)\right]d\Sigma^2.
	\end{split}
\end{equation}
Following the same procedure as before, it is found that $\beta_A=-\beta_R$. We see that the variation of solution can be written as 
\begin{equation}
	\delta_V g_{\mu\nu}dx^\mu dx^\nu=\ell\beta_R\sin\left(\frac{2\tau}{\ell}\right)d\Sigma^2,
\end{equation}
a transformation which is generated by the vector field $Y=-\,\beta_R \partial_{\tau}$, \textit{i.e.} $\delta_V g_{\mu\nu}=\mathcal{L}_Y g_{\mu\nu}$.

To read off the boundary data, we set $\ell=1$, take $d\Sigma^2=(du^2+d\vec x^2)/u^2$ and pass to Poincar\'e coordinates through $t=u\tan\tau$, $z=u\sec\tau$, in which the metric becomes $(-dt^2+dz^2+d\vec x^2)/z^2$ \cite{Carrasco:2023fcj}. After continuing to Euclidean signature, $t\to-it_E$, the deformation $\mathcal L_Y g$ is not in Fefferman-Graham gauge. We restore the gauge order by order in $z$ in Appendix~\ref{app:FG_York}. Near the boundary, the combined transformation acts as the radial rescaling generated by $Y_{\rm bdry}=i\beta_R\,z\,{\rm sign}(t_E)\,\partial_z$, in agreement with~\cite{Carrasco:2023fcj}. Hence
\begin{equation}
	\delta_V\gamma_{ab}=-2i\beta_R\,{\rm sign}(t_E)\,\gamma_{ab},\qquad \delta_V\braket{T_{ab}}=0,
\end{equation}
\textit{i.e.} the York deformation acts on the boundary as a Weyl transformation, constant on each half, and leaves the stress tensor unchanged up to the anomalous contribution, which is absent for odd $d$ \cite{deHaro:2000vlm}.

Plugging this result into the definition of the boundary symplectic form we find that 
\begin{equation} \label{eq:VolumeOmegaBdry}
	\begin{split}
		\Omega_{\partial}(\delta,\delta_V)&=\frac{i}{2}\int_{\mathcal{B}_+}d^d x\sqrt{\gamma}\left(\delta \gamma_{ab}^+\delta_V^- \braket{T^{ab}}-\delta_V\gamma_{ab}^+\delta^- \braket{T^{ab}}\right)\\
		&=-\beta_R\int_{\mathcal{B}_+}d^d x\sqrt{\gamma}\delta^-\braket{T_{a}^a}.
	\end{split}
\end{equation}

Now we combine \eqref{eq:deltaV}, \eqref{eq:VolumeOmegaBdry} using \eqref{eq:transport_symplectic_pairing}, where now $\delta_V$ represents its boundary representation, to write
\begin{equation}
	\frac{\delta V}{G_N \ell} = \Omega_{\mathrm{bulk}}(\delta,\delta_V) = \Omega_{\partial}(\delta,\delta_V)=-\beta_R\int_{\mathcal{B}_+}d^d x\sqrt{\gamma}\delta^-\braket{T_{a}^a},
\end{equation}

By the Ward identity, the trace $\braket{T^a_{\phantom{a}a}}$ is equal to the Weyl anomaly, which is a local function of the sources. Its variation with respect to the sources localised on $\mathcal B_-$ hence vanishes at the upper-half points separated from the deformation cut, where the Weyl transformation induced by the York deformation jumps. Introducing a cut-off near $\tau=0$ and ignoring corner contributions, $\Omega_\partial(\delta,\delta_V)=0$. This relation need not hold around a different reference state, such as black holes, where applying the Peierls construction is notoriously harder.

In order to find a non-trivial source-dependent contribution to the boundary generator, we are going to introduce a scalar field into the picture\footnote{The bulk is the same AdS$_{d+1}$ in both cases but the conformal frame differs. In Poincaré coordinates, the boundary is flat, while in WdW coordinates each half $\mathcal{B}_\pm$ carries the hyperbolic metric, which is the flat one rescaled by $t_E^{-2}$ since $t_E=\pm u$ at the boundary. $\Omega_\partial$ and the trace Ward identity are Weyl invariant up to local terms, which do not contribute to $\Omega_\partial$. Then, we can read $\delta_Y$ in the flat frame and the scalar sources in the WdW frame.}.

\subsection{Adding a scalar field}\label{sec3.1:scalar_field}
In this subsection, we will repeat the former calculation but in the presence of a scalar field with source $\lambda$ and conformal dimension $\Delta$. For simplicity, we will consider $d=2$ and $\ell=1$. In addition to the Einstein-Hilbert term, the action includes the piece for the scalar field, and reads
\begin{equation}
	S=\frac{1}{16\pi G_N}\int_M d^3 x \sqrt{-g} (R-2\Lambda)-\frac{1}{2}\int_M d^3x \sqrt{-g}(g^{\mu\nu}\nabla_\mu\phi \nabla_\nu \phi+m^2\phi^2) +S_{GHY}.
\end{equation}
The observable here is simply
\begin{equation}
	V=\frac{1}{G_N \ell}\int_{\Sigma}d^2 \sigma \sqrt{h},
\end{equation}
where $\Sigma$ is the maximal slice anchored at $\Sigma_{\rm CFT}$, which can be taken to be the $\tau=0$ slice at the order we work (see below) and $\sigma$ represent the coordinates on this hypersurface. After computing the variation of the ``perturbed'' action, the equations of motion are found
\begin{equation}
	\begin{split}
		\left(\Box-m^2\right)\phi &=0,\\
		R_{\mu\nu}-\frac{1}{2}R g_{\mu\nu}+\Lambda g_{\mu\nu}+8\pi\epsilon\,h_{\mu\nu}\delta(\tau)
		&=8\pi G_N T^{\text{scalar}}_{\mu\nu},
	\end{split}
\end{equation}
with $T^{\text{scalar}}_{\mu\nu}$ is the stress energy tensor associated to the scalar sector. 

As we work at quadratic order in the sources, two simplifications follow, since the volume changes at order $O(\lambda^2)$ and its variation at $O(\lambda\delta\lambda)$. First, the gravitational  contribution to the symplectic pairing enters at this same order but, as shown below, the combined gravitational and scalar contribution of the diffeomorphism part of the York deformation vanishes by the boundary Ward identity in the regulated pairing. The remaining contribution can therefore be evaluated with the scalar field and its compensating deformation at $O(\lambda)$ on the AdS background. Second, since the metric differs from AdS by $O(\lambda^2)$, the first variation of the volume at $\tau=0$ is $O(\lambda^2)$, while the second variation is $O(\lambda^0)$ and negative definite, as in the vacuum. The maximal slice is therefore displaced by $O(\lambda^2)$, which changes the volume only at $O(\lambda^4)$, and being extremal, its displacement does not contribute to $\delta V$. We may thus take $\Sigma=\{\tau=0\}$ and hold it fixed in the Peierls construction.

As the volume deformation is localised at $\tau=0$, the solutions for $\tau>0$ and $\tau<0$ are homogeneous, so we only have to focus on matching them. As the volume functional does not explicitly depend on the scalar field, there is no jump in either $\varphi$ or $\pi_\phi$. On the other hand, for the retarded solution, the induced metric $h_{ij}$ and extrinsic curvature tensor $K_{ij}$ have a discontinuity given by \eqref{eq:retarded_constraints}. The advanced solution satisfies a similar relation but with opposite sign.

Thus, the Peierls variation generated by the volume obeys
\begin{equation}
	\delta_VK_{\mu\nu}=8 \pi h_{\mu\nu},     \qquad
	\delta_Vh_{\mu\nu}
	=\delta_V\varphi     =\delta_V\pi_\phi=0
	\quad\text{on }\Sigma.
\end{equation}
From the previous example, it is known that the vector field $Y=-8\pi \partial_\tau$ generates the appropriate variation of the induced metric and extrinsic curvature tensor. Nevertheless, as it acts as a diffeomorphism, it also gives rise to a non-trivial variation on the field $\varphi$ and its conjugate momentum $\pi_\phi$. For this reason, we decompose the variation of the fields as
\begin{equation}
	\begin{split}
		\delta_V g_{\mu\nu}&=\mathcal{L}_{Y}g_{\mu\nu},\\
		\delta_V\phi
		&=\mathcal{L}_{Y}\phi+\delta'\phi,
	\end{split}
\end{equation}
where $\delta'\phi$ is chosen so that the total deformation leaves the scalar initial data unchanged, $\delta_V\varphi|_\Sigma=\delta_V\pi_\phi|_\Sigma=0$. For this reason, 
\begin{equation}\label{eq: constraint delta prime}
	\delta'\varphi= \delta'\phi|_\Sigma
	=8\pi\,\partial_\tau\phi|_\Sigma,     \qquad
	\delta'\pi_\phi|_\Sigma
	=\left.8\pi \partial_\tau\pi_\phi\right|_\Sigma=\left.8\pi\sqrt{h}\, \partial_\tau^2\phi\right|_\Sigma.
\end{equation}
Moreover, the paired deformation
$(\mathcal L_Y g,\mathcal L_Y\phi)$ solves the coupled
linearised equations. Since the metric part of the York deformation agrees with $\mathcal L_Y g$ at the order
considered, the compensating deformation $\delta'\phi$ obeys the homogeneous Klein-Gordon equation on the zeroth-order AdS background. We will focus on the case in which the scalar field has a large-conformal dimension $\Delta$ since in that limit the source integral simplifies. In Section~\ref{sec:finite_dimension_cost} we treat the general case. Henceforth, we will work in WdW coordinates in Euclidean signature where the future boundary $\mathcal{B}_+$ is located at $\tau_E\rightarrow \infty$. The boundary-to-bulk propagator is
\begin{equation} \label{eq:boundary_to_bulk_propagator}
	K_E(\tau_E,u,x|u_0,x_0)=\frac{\Delta-1}{4\pi}\left(\frac{1}{\cosh \tau_E}\frac{2u u_0}{(x-x_0)^2+u^2+u_0^2-2su u_0\tanh \tau_E}\right)^{\Delta}
\end{equation}
where $s=1$ when $(u_0,x_0)$ is in the top boundary and $s=-1$ when it is in the bottom boundary. To find the solution when sources $\lambda^+,\lambda^-$ are turned on over $\mathcal{B}_+$ and $\mathcal{B}_-$ respectively, we just have to integrate this propagator against these sources. Assuming that $\Delta$ is large and that $\lambda^\pm$ are translation invariant along $x$, the integral yields
\begin{equation}
	\begin{split} 
		\phi(\tau_E)=&\int_{\partial\mathcal{M}} \frac{dx_0 du_0}{u_0^2}K_E(\tau_E,u,x|u_0,x_0)\lambda(u_0)\\
		&\approx\frac{\Delta-1 }{2\Delta \cosh \tau_E}\left(e^{(1-\Delta)\tau_E}\lambda^-(u)+e^{(\Delta-1)\tau_E}\lambda^+(u)\right) \label{eq:source-int}
	\end{split}
\end{equation}
This field generates the following Lorentzian initial data on the $\tau_E=0$ slice\footnote{Lorentzian initial data are related to Euclidean data by $\phi|_L=\phi|_E$ and $\pi_\phi|_L=i\pi_\phi|_E$}
\begin{equation}
	\begin{split}
		\varphi
		&\simeq\frac{1}{2}
		\left(\lambda^-(u)+\lambda^+(u)\right),\\
		\pi_\phi
		&\simeq\frac{i\Delta}{2u^2}
		\left(\lambda^+(u)-\lambda^-(u)\right).
	\end{split}
\end{equation}
Taking into account the variation of these initial data and the constraint in \eqref{eq: constraint delta prime}, we can determine which is the variation $\delta'$ of the sources
\begin{equation}
	\delta'\lambda^+(u)
	=8 i\pi \Delta\,\lambda^+(u),
	\qquad
	\delta'\lambda^-(u)
	=-8 i\pi \Delta\,\lambda^-(u).
\end{equation}
Now that we have the variation of the sources, let us focus on the boundary symplectic form. We stress that this form consists of two different pieces: one involving the diffeomorphism generated by the York vector $Y$; and another coming from $\delta'\phi$. Remember that $Y$ acts on the boundary as a Weyl transformation, which rescales all boundary data. In particular, the boundary metric (in Poincaré coordinates) is rescaled by a factor of $(1-16i\pi\epsilon\ell \textup{ sign}(t_E))$ which comes from a rescaling of the radial coordinate $z$ of the form $z=(1+8i\pi\epsilon \text{ sign(}t_E))z'$. If the scalar field has conformal dimension $\Delta$ and is expanded as a power series in $z$, the leading term would be of the form $\phi=z^{d-\Delta }\lambda$. After the aforementioned rescaling, the scalar field would read 
\begin{equation}
	\phi= z'^{(d-\Delta)} e^{8\pi i \epsilon(d-\Delta)\textup{ sign}(t_E)}\lambda\approx z'^{d-\Delta}(1+8\pi i \epsilon(d-\Delta)\textup{sign}(t_E))\lambda
\end{equation}
which gives $\delta_V\lambda=8\pi i(d-\Delta)\,\textup{sign}(t_E)\,\lambda$. A similar analysis gives $\delta_V\braket{\mathcal O}=8\pi i\Delta\,\textup{sign}(t_E)\braket{\mathcal O}$. Once this variation is plugged into the symplectic form it is found 
\begin{equation}
	\Omega_\partial(\delta, \delta_V)=-8\pi\int_{\mathcal B_+}d^dx\,\sqrt{\gamma}\,
	\delta^-\left[
	\braket{T^a{}_a}
	+(d-\Delta)\lambda^+\braket{\mathcal O} \right].
\end{equation}
By the trace Ward identity, $\braket{T^a_a}$ equals the Weyl anomaly, a local functional of the sources. Its variation with respect to lower-half sources vanishes at upper-half points separated from the preparation cut. Thus, with the cut regulated and assuming no remaining corner contribution, this pairing vanishes. The remaining contribution therefore comes from the compensating scalar deformation $\delta'\phi$.
\begin{equation}
	\begin{split}
		\Omega_{\rm bulk}(\delta,\delta')&\simeq\int_\Sigma du\,dx\,
		\left(\delta\pi_\phi\,\delta'\varphi-\delta'\pi_\phi\,\delta\varphi \right)\\
		&\simeq 4\pi \Delta^2\int\frac{du\,dx}{u^2}
		\left( \lambda^+ \delta\lambda^-
		+\lambda^-\delta\lambda^+
		\right)\\
		&=\delta\left(
		4\pi  \Delta^2\int\frac{du\,dx}{u^2}
		\lambda^-\lambda^+
		\right).
	\end{split}
\end{equation}

Here, we evaluated the pairing on $\Sigma=\{\tau=0\}$. This is the slice along which the Euclidean saddle is cut, \textit{i.e.} the boundary of $\mathcal M_+$ in Section~\ref{subsec:fs_boundary_symplectic}, so the argument of~\cite{Belin:2018fxe} reviewed there applies directly. Both $\delta$ and $\delta'$ are linearised solutions generated by source variations, $d\omega_E=0$ moves the pairing from $\mathcal{B}_+$ to $\Sigma$ and the continuation $\pi_L=i\pi_E$ turns it into~\eqref{eq:symplectic_form_scalar}. Hence, $\Omega_{\rm bulk}(\delta,\delta')=\Omega_{\partial}(\delta,\delta')$, and within the small-source, large-$\Delta$ approximation, the transported volume generator is a quadratic functional of the preparation sources, determined up to an additive constant, providing a concrete example for the local preparation-cost construction developed in the next section.

\section{From transported flows to boundary cost geometry}\label{sec4: flow_to_cost}

We have established a prescription through which we can find boundary flows generated by bulk observables. These can in turn be related to boundary observables. As a proof of principle, we described in detail the volume observable in the main text and the extremal area in Appendix~\ref{app:B}. Now, starting from the boundary generator $H_W^{\rm CFT}$ associated with a bulk observable $W_\textup{bulk}$, we define a metric in the space of sources, locally near the reference state whose associated notion of complexity in the CFT side satisfies a first law of complexity of the form $\delta \mathcal{C}=\delta H_W^{\rm CFT}=\delta W_{\text{bulk}}$. 

Nielsen's approach recasts circuit complexity as a variational problem \cite{Nielsen:2005mkt,Nielsen:2006cea,Dowling:2006tnk, Brown:2017jil}. This idea has been extended to quantum field theories \cite{Jefferson:2017sdb,Chapman:2017rqy}. Here we adopt the version that  suits best the holographic setup established in Section \ref{sec2:holographic_transport}, where the family of states prepared via insertion of Euclidean sources is regarded as a manifold $\mathcal{S}$ coordinatised with the sources $\lambda^a$. A circuit can be regarded as a path $\lambda(s)$ in this manifold joining the reference $\lambda_i$ and target state $\lambda_f$. To assign a weight to each possible curve connecting them, the so-called \textit{cost function} $F(\lambda, \dot{\lambda})$ is defined and the complexity of constructing the target state from the reference state is obtained after extremising the functional
\begin{equation}
    \mathcal{C}(\lambda_i,\lambda_f)=\min_{\lambda(s)}\int_{s_i}^{s_f}ds F(\lambda,\dot{\lambda}).
\end{equation}

This scheme requires us to know what the cost function is. Evidence has been gathered that, for the bulk volume, the relevant cost is a kinetic energy for the sources \cite{Belin:2018bpg}. Related proposals assign costs to holographic path integrals through bulk functionals \cite{Chandra:2022pgl}. In general, however, there is no prescription that singles out a specific form for this function starting from a given bulk observable.

A related ambiguity comes precisely from these holographic perspectives where many observables in the gravitational side can take the role of duals to complexity \cite{Belin:2021bga,Belin:2022xmt,Jorstad:2023kmq,Myers:2024vve}. Although we do not claim to have solved either ambiguity, it is interesting to study how the two are related in more detail. So equipped with our holographic transport scheme presented above we now proceed to make a more precise, albeit conjectural, statement about how these might be related: a bulk observable satisfying suitable conditions determines a particular metric in a neighbourhood of the reference state such that the ``kinetic energy'' integral between the reference and target state gives a notion of complexity which equals the transported boundary generator to quadratic order.

\subsection{An observable-adapted metric}\label{sec4.1: observable_metric}
Let $W_{\textup{bulk}}$ be a bulk observable and $H_W^{\rm CFT}$ the dual boundary generator obtained via the procedure described in Section \ref{sec2:holographic_transport}. Two assumptions will be considered for this observable: first, the reference state is a critical point for this observable, $\delta H_W^{\rm CFT}|_{\lambda_i}=0$ and second, the Hessian of this observable is positive definite at the same state. These two conditions are motivated as follows. Complexity of the reference state vanishes by definition, and it must be positive for any other state. Then, if it is to coincide with $H_W^{\rm CFT}-H_W^{\rm CFT}(\lambda_i)$ it must be a local minimum of $H_W$, which requires these two assumptions\footnote{Actually, to be a local minimum requires the Hessian matrix to be positive semi-definite. We assume in addition that the Hessian is non-degenerate and hence, positive definite.}.

We denote by $\delta_W$ the transported flow, which verifies
\begin{equation}
    \delta H_W^{\rm CFT}=\Omega_\partial(\delta,\delta_W),\quad \forall \delta \in T\mathcal{S}
\end{equation}
Hamiltonian vector fields are defined up to zero modes of $\Omega_\partial$, and the construction below applies to any subfamily of sources on which $\Omega_\partial$ is non-degenerate and the Hessian of $H_W^{\rm CFT}$ is positive definite.
In sources coordinates and writing $\lambda=\lambda_i+v$, we expand
\begin{equation}
    H_W^{\rm CFT}(\lambda)=H_W^{\rm CFT}(\lambda_i)+\frac{1}{2}\mathcal{H}(v,v) +\mathcal{O}(v^3)
\end{equation}
where the symmetric bilinear form $\mathcal{H}(u,v)$ is the aforementioned Hessian and is defined as
\begin{equation}
    \mathcal{H}(u,v)=\sum_{a,b}\int \int d^dx d^dy \, u^a(x)\left.\frac{\delta^2 H_W^{\rm CFT}}{\delta\lambda^a(x)\delta\lambda^b(y)}\right|_{\lambda_i}v^b(y)
\end{equation}
Here, $a,b$ run over all the sources and their conjugates which, as in Section \ref{sec2:holographic_transport}, are treated as independent variables. Thus, $\mathcal{H}$ is a real bilinear form once the reflection condition is imposed. Since $\delta H_W$ vanishes at $\lambda_i$ and the symplectic form is non-degenerate, the flow induced by the boundary generator is also vanishing. Its linearisation defines a map $A:T_{\lambda_i}\mathcal{S}\rightarrow T_{\lambda_i}\mathcal{S}$ such that
\begin{equation}
    \delta_W|_{\lambda_i+v}=A v+O(v^2),\quad \Omega_\partial( u,Av)=\mathcal{H}(v,u)
\end{equation}
This map meets the following properties:
\begin{enumerate}
    \item  $A$ is antisymmetric with respect to the symplectic form $\Omega_\partial$. Indeed, from the definition of $A$ and the symmetry of $\mathcal{H}$ 
    $$\Omega_{\partial}(u,Av)=\mathcal{H}(v,u)=\mathcal{H}(u,v)=\Omega_{\partial}(v,Au)=-\Omega_{\partial}(Au,v).$$
    \item $A$ is antisymmetric with respect to $\mathcal{H}$.
    $$\mathcal{H}(Au,v)=\Omega_{\partial}(v,A^2 u)=-\Omega_{\partial}(Av,Au)=-\mathcal{H}(Av,u).$$
    \item $A$ is an invertible map. If $A u=0$ then $\mathcal{H}(u,v)=\Omega_\partial(v,Au)=0$ for all $v$ and, as $\mathcal{H}$ is non-degenerate, this implies $u=0$. 
    \item The eigenvalues of $A$ are strictly imaginary. Since $\mathcal{H}$ is positive definite, we can define an inner product as $\braket{u,v}=\mathcal{H}(u^*,v)$. Let $\alpha$ be an eigenvector of $A$ with eigenvalue $\mu$. Then,
    $$\mu\braket{\alpha,\alpha}=\mathcal{H}(\alpha^*,A\alpha)=-\mathcal{H}(A\alpha^*,\alpha)=-\mu^*\braket{\alpha,\alpha}$$
    From the previous expression, we deduce that $\mu=-\bar{\mu}$ which can only happen if $\mu$ is purely imaginary\footnote{The case $\mu=0$ is discarded because $A$ is invertible.}. The eigenvectors will be labeled as
    $$A\alpha_k=i\omega_k \alpha_k,\quad A\alpha^*_k=-i\omega_k\alpha^*_k$$
\end{enumerate}
Since $A$ is antisymmetric with respect to $\mathcal{H}$, $-A^2$ is positive. We define the operator $|A|\equiv \sqrt{-A^2}$, which by definition, commutes with $A$ and satisfies $|A|\alpha_k=\omega_k\alpha_k$. The map $A$ determines a complex structure $J_W$ as follows \cite{daSilva2008}
\begin{equation}
    J_W=A|A|^{-1},\quad J_W\alpha_k=i\alpha_k,\quad J_W\alpha_k^*=-i\alpha^*_k.
\end{equation}
One easily checks that $J_W$ is a complex structure satisfying $J_W^2=-1$, $\Omega_\partial(J_Wu, J_W v)=\Omega_\partial(u,v)$ and the bilinear form
\begin{equation}\label{eq: Kahler_metric}
    g_J(u,v)\equiv\Omega_{\partial}(u,J_W v)=\mathcal{H}\left(u,|A|^{-1}v\right)
\end{equation}
is symmetric and definite positive. Hence, $J_W$ is compatible with our symplectic form $\Omega_\partial$. In addition, it commutes with $A$ by construction, and it is the only complex structure that satisfies all these four properties. Note that the metric $g_J$, unless $A$ has a single eigenvalue $\omega$, is not proportional to the Hessian $\mathcal{H}$. 

Finally, we define a cost metric directly from the Hessian $\mathcal{H}$ 
\begin{equation}\label{eq: hessian_metric}
    G(u,v)= \frac{1}{2} \mathcal{H}(u,v).
\end{equation}
and a notion of complexity through the energy functional as
\begin{equation}
    \mathcal{C}(\lambda_i,\lambda_f)=(s_f-s_i)\int_{s_i}^{s_f} G(\dot{\lambda}_*,\dot{\lambda}_*)ds
\end{equation}
with $\lambda_*$ the extremal curve joining the reference and target state.

Notice this metric has constant components in the source coordinates and, hence, its geodesics are straight lines. The optimal curve connecting both the reference and target state is $\lambda_*(s)=\lambda_i+\frac{s-s_i}{s_f-s_i}v$ with $v=\lambda_f-\lambda_i$. Therefore,
\begin{equation}
    \mathcal{C}(\lambda_i,\lambda_f)=G(v,v)=\frac{1}{2}\mathcal{H}(v,v)=H_W^{\rm CFT}(\lambda_f)-H_{W}^{\rm CFT}(\lambda_i)+O(v^3).
\end{equation}
Among metrics with constant components in source coordinates, $G=\frac{1}{2}\mathcal H$ is the only one for which the minimised energy reproduces $H_W^{\rm CFT}$ to quadratic order, since requiring $G(v,v)=\frac{1}{2}\mathcal H(v,v)$ for all $v$ fixes $G$ by polarisation.

The first law of complexity follows directly from the variation of the energy functional, as explained in \cite{Bernamonti:2019zyy,Bernamonti:2020bcf}. Fixing the reference and the parameter range, the variation of complexity reduces to the variation of the bulk observable.
\begin{equation}
    \delta\mathcal{C}=2(s_f-s_i) G(\dot \lambda_*(s_f),\delta\lambda_f)=\mathcal{H}(v, \delta \lambda_f)=\Omega_\partial(\delta \lambda_f,\delta_W)=\delta H_W^{\rm CFT}=\delta W_{\textup{bulk}},
\end{equation}
result valid up to linear order in $v$. 

It is instructive to express the terminal velocity of the optimal curve in terms of the transported flow and the complex structure. Given that $A^{-1}=-|A|^{-1}J_W$,
\begin{equation}
    \dot{\lambda}_*(s_f)=\frac{v}{s_f-s_i},\quad v=\left.A^{-1}\delta_W\right|_{\lambda_f}+O(v^2)=-\left.|A|^{-1}J_W \delta_W\right|_{\lambda_f}+O(v^2).
\end{equation}
From this expression we deduce that the optimal curve does not follow the flow generated by $H_W$ if it were treated as a Hamiltonian function but it is the transported flow rotated by the complex structure $J_W$ and then rescaled by the map $|A|^{-1}$. To leading order $-|A|^{-1}J_W \delta_W$ is the radial field centred at the reference state, whose integral lines are the optimal straight lines.

The content of the proposal lies in the choice of preparation penalties, fixed by the observables via the Hessian $\mathcal H$ and in the relation between optimal circuits and the transported flow. To quantify them, note that the cost metric $G$ is a quadratic form on $T\mathcal S$. How penalised a direction is only makes sense relative to a notion of how large it is. We have to establish first which directions are equally large and then compare their cost. The states themselves provide that notion. We regard two directions as equally large if moving along them by the same amount produces states that are equally distinguishable  from the reference state. For a displacement $s\delta$, the fidelity with the reference state is~\cite{Zanardi:2007byf}
\begin{equation}
     \frac{|\langle\psi_{\lambda_i}|\psi_{\lambda_i+s\delta}\rangle|^2}{\langle\psi_{\lambda_i}|\psi_{\lambda_i}\rangle\,\langle\psi_{\lambda_i+s\delta}|\psi_{\lambda_i+s\delta}\rangle}
    =1-s^2\,g_{\rm FS}(\delta,\delta)+O(s^3),
\end{equation}
where $g_{\rm FS}$ is the Fubini-Study metric in~\eqref{eq:fs_metric_general}. Directions of equal Fubini-Study length are equally large in this sense so we take this metric as the natural reference. This information is not contained  in the symplectic form alone. For instance, a transformation acting on a pair of canonical coordinates as $(q,p)\rightarrow(e^r q,e^{-r}p)$ preserves $\Omega_\partial$ but changes the size of a displacement of $q$ relative to one of $p$. The symplectic form thus cannot tell whether a field displacement is larger or smaller than a momentum displacement, whereas the inner product can\footnote{Equivalently, $g_{\rm FS}$ is the only metric on $T_{\lambda_i}\mathcal S$ invariant under the linear transformations preserving both $\Omega_\partial$ and $J_{\rm FS}$. These act transitively on directions of unit length, so no direction is privileged. Uniqueness follows because any invariant symmetric bilinear form is $g_{\rm FS}(\,\cdot\,,\,T\cdot\,)$ with $T$ commuting with the group, \textit{i.e.} $T=a+b J_{\rm FS}$. Symmetry of this metric requires $b=0$ since $J_{\rm FS}$ is antisymmetric with respect to $g_{\rm FS}$.}. The Fubini-Study metric plays the role of an unpenalised metric in Nielsen's approach. On the full projective Hilbert space, it is the metric induced by the bi-invariant metric on the unitary group.

We can quantify how expensive a direction is by comparing its cost with its Fubini-Study length. We define the penalty of a direction $\delta$ as the ratio
\begin{equation}\label{eq:penalty_definition}
    p(\delta)=\frac{G(\delta,\delta)}{g_{\rm FS}(\delta,\delta)}.
\end{equation}
Penalty factors of this kind are generally introduced by hand according to Nielsen's approach, including CFTs \cite{Baiguera:2025uss}. Here they are fixed by the bulk observable through the Hessian.

Directions with a larger penalty are more expensive to prepare than what their Fubini-Study length indicates. 

Generally, $p(\delta)$ depends on the direction and a natural way to characterise this dependence is through its stationary values. Provided that both $G$ and $g_{\rm FS}$ are symmetric and positive definite, one can solve the eigenvalue problem $\det(G-p\,g_{\rm FS})=0$, which yields a set of principal directions $\tilde\delta_i$ and real eigenvalues $p_i$. In this basis, it is verified
\begin{equation}\label{eq:principal_penalties}
    G(\,\cdot\,,\tilde \delta_i)=p_i\, g_{\rm FS}(\,\cdot\,,\tilde\delta_i).
\end{equation}
We call $p_i$ the principal penalties. Note that~\eqref{eq:principal_penalties} does not state that $G$ and $g_{\rm FS}$ are proportional but it holds only along the principal directions.

Since $G(\delta_1,\delta_2)=\frac{1}{2}\Omega_\partial(\delta_1,A\delta_2)$ and $\Omega_\partial$ is non-degenerate~\eqref{eq:principal_penalties} is equivalent to $A\tilde\delta_i=p_i J_{\rm FS}\tilde\delta_i$ \textit{i.e.} 
\begin{equation}\label{eq:penalty_operator}
    -J_{\rm FS} A \tilde\delta_i=p_i\tilde\delta_i
\end{equation}
The principal penalties are then the eigenvalues of $-J_{\rm FS}A$.

\subsection{Check with volume}\label{subsec: check_volume}
To test the validity of this proposal, we will apply it to the complexity=volume proposal in the setup described in Section \ref{sec3: CV} in the presence of a scalar field with large-conformal dimension. We choose this setup because, unlike the pure gravity case, the boundary observable dual to volume does not vanish identically.

Let us recall that the symplectic form evaluated at the York deformation is
\begin{equation}\label{eq:delta U}
\begin{split}
    \Omega_\partial(\delta,\delta_V)=\delta\left(4\pi  \Delta^2\int \frac{du\; dx}{u^2}  \lambda^-(u)\lambda^+(u)\right),
\end{split}
\end{equation}
Then, the boundary observable reads
\begin{equation}
    H_V^{(2)}=4\pi \Delta^2  \int \frac{du\; dx}{u^2}\lambda^-(u)\lambda^+(u).
\end{equation}
This observable is quadratic in the sources, which means that the reference state $\lambda^+=\lambda^-=0$, is a critical point and $H_V^{(2)}=\frac{1}{2}\mathcal{H}$ exactly within this approximation. Once the reflection condition \eqref{eq:reality_condtion} is imposed, $H_V^{(2)}= 4\pi \Delta^2\int \frac{du\;dx}{u^2}|\lambda^-|^2$ yields a positive definite Hessian, so the conditions imposed in Section \ref{sec4.1: observable_metric} are satisfied.

Since $\Omega_\partial(\delta,\delta_Y)=0$ for all $\delta$ due to the trace Ward identity, the York diffeomorphism is a zero mode of $\Omega_\partial$ and the Hamiltonian vector field of $H_V^{(2)}$, defined up to zero modes (Appendix~\ref{app1:Review}), can be taken to be $\delta'$. As the Hessian is degenerate along metric variations and does not mix them with scalar ones, we restrict to scalar sources at fixed boundary metric. There, $\Omega_\partial$ is weakly non-degenerate and $A$ is the linearisation of $\delta'$.

To determine the complex structure $J_V$ associated to this observable, we start with the boundary symplectic form of Section \ref{sec3.1:scalar_field} evaluated on the large-$\Delta$ data of Section~\ref{sec3.1:scalar_field}
\begin{equation}
    \Omega_\partial(\delta_1,\delta_2)=\frac{i\Delta}{2}\int \frac{du\; dx}{u^2}(\delta \lambda^+_1\delta\lambda^-_2-\delta\lambda^+_2\delta\lambda^-_1).
\end{equation}
Comparing this expression with \eqref{eq:delta U}, we read off the action of the transported flow on the sources
\begin{equation}
    \delta' \lambda^+=8i \pi \Delta\lambda^+,\quad \delta' \lambda^-=-8i \pi \Delta\lambda^-.
\end{equation}

Since the flow is linear in the sources, the map $A$ acts in the same way in the sources. The flow is then a phase rotation of the sources $\lambda^\pm\rightarrow e^{\pm 8\pi i \Delta s}\lambda^\pm$, which preserves the reflection condition. As we can see, all the frequencies $\omega$ are the same and $|A|=8\pi \Delta\mathbf{1}$. The complex structure acts on the sources as
\begin{equation}
    J_V \lambda^+=i\lambda^+,\quad J_V\lambda^-=-i\lambda^-.
\end{equation}
This coincides with the complex structure compatible with both the symplectic form $\Omega_\partial$ and the Fubini-Study metric $g_{FS}$ introduced in Section \ref{subsec:fs_boundary_symplectic}. This result agrees with that of \cite{Belin:2018bpg,Belin:2018fxe} up to an overall sign coming from the opposite sign convention for the symplectic form. Since the frequency spectrum is degenerate and there is a single frequency, the metric \eqref{eq: hessian_metric} is proportional to  the Kähler metric \eqref{eq: Kahler_metric}. More concretely,
\begin{equation}
    G(\delta_1,\delta_2)=\frac{1}{2}\mathcal{H}(\delta_1,\delta_2)=2\pi \Delta^2\int \frac{du\; dx}{u^2}(\delta_1 \lambda^+\delta_2\lambda^-+\delta_2 \lambda^+\delta_1\lambda^-)
\end{equation}
while
\begin{equation}
    g_{J}(\delta_1,\delta_2)= \frac{\Delta}{2}\int \frac{du\,dx}{u^2}(\delta_1 \lambda^-\delta_2\lambda^++\delta_2\lambda^-\delta_1\lambda^+),
\end{equation}
in such a way that $G(\delta_1,\delta_2)=4\pi \Delta\, g_{J}(\delta_1,\delta_2)$. With the normalisation $g_{FS}=\frac{1}{2}\Omega_\partial(\cdot,J_V\cdot)$ of Section \ref{subsec:fs_boundary_symplectic} this gives $G=8\pi \Delta g_{\rm FS}$, recovering the proportionality to the Fubini-Study metric found in \cite{Belin:2018bpg} through a different route. In the language of Section~\ref{sec4.1: observable_metric}, all directions have the same penalty, $p(\delta)=8\pi \Delta$ and the volume complex structure coincides with $J_{\rm FS}$ and the transported flow has a single frequency.

\subsection{Finite conformal dimension and preparation penalties}
\label{sec:finite_dimension_cost}

In Section \ref{subsec: check_volume} we found that, at leading order in large-conformal dimension, all directions have the same penalty, and the cost metric is proportional to the Fubini-Study metric. We now ask whether this proportionality still holds when the full dependence on the conformal dimension $\Delta$ is retained. We keep the setup of Section~\ref{sec3.1:scalar_field} with arbitrary $\Delta\geq 2$ and work at quadratic order around the AdS vacuum. The sources are assumed to vanish sufficiently fast near the preparation cut and as $u\rightarrow 0,\infty$ so that the integration by parts below have no boundary contribution. Here we determine the principal penalties and the complex structure selected by the volume.

As in Section \ref{sec3.1:scalar_field}, the volume flow consists of the York deformation and a compensating scalar deformation $\delta'$, fixed by~\eqref{eq: constraint delta prime} so that the total flow leaves the scalar initial data unchanged. On $\Sigma=\{\tau=0\}$  the Klein-Gordon equation reads $\partial_\tau^2\phi|_{\Sigma}=(\nabla^2_{\mathbb{H}^2}-m^2)\varphi$, with $m^2=\Delta(\Delta-2)$ and $\nabla^2_{\mathbb{H}^2}$ is the Laplacian of the slice $\Sigma$. Hence,  

\begin{equation}
    \delta'\varphi=8\pi \frac{\pi_\phi}{\sqrt{h}},\qquad\delta'\pi_\phi=8\pi\sqrt h \left(\nabla_{\mathbb{H}^2}^2-m^2\right)\varphi
\end{equation}

As happened above, the diffeomorphism contribution cancels in the regulated pairing, so the generator is determined by the scalar pairing. Inserting $\delta'$ in~\eqref{eq:symplectic_form_scalar} and integrating by parts gives 
\begin{equation}\label{eq:finite_delta_energy_pairing}
    \Omega_\partial(\delta,\delta')=\delta H_V^{(2)},\quad \text{with}\quad H_V ^{(2)}=4\pi\int_\Sigma d^2 x\left[\frac{\pi_\phi^2}{\sqrt h}+\sqrt{h}\left(h^{ij}\partial_i\varphi\partial_j\varphi+m^2 \varphi^2\right) \right].
\end{equation}
up to an additive constant, where $h_{ij}$ is the metric induced on $\Sigma$. Variations of the metric in this expression contribute at order $O(\lambda^3\delta\lambda)$, so here they are neglected.

If we assume $x-$independent configurations, $\sqrt{h}=1/u^2$ and $\nabla_{\mathbb{H}^2}^2=u^2\partial_u^2$. The latter operator is self-adjoint with respect to $du/u^2$ and has the following eigenfunctions
\begin{equation}\label{eq:basis}
    e_{k}=\frac{1}{\sqrt{2\pi}} u^{1/2+i k},
\end{equation}
which are orthonormal with respect to the measure $du/u^2$ and have the eigenvalue $-\sigma_k=-(1/4+k^2)$. Expanding  $\varphi=\int dk\,\varphi_k\,e_k$ and $u^2\pi_\phi=\int dk\, \pi_k\,e_k$,~\eqref{eq:finite_delta_energy_pairing} becomes, per unit of $x-$length
\begin{equation}\label{eq:finite_delta_generator}
    H_V^{(2)}=4\pi\int dk\left[\pi_k\pi_{-k}+\left(\sigma_k+m^2\right)\varphi_k\varphi_{-k}\right].
\end{equation}
We now express $\varphi_k$ and $\pi_k$ in terms of the sources by evaluating the source integral~\eqref{eq:source-int}, but without assuming the large-$\Delta$ approximation. A source mode $u_0^{1/2+i k}$ on either half then produces a bulk profile given by the integral of $K(0,u,x|u_0,x_0)$ over $u_0$ and $x_0$, that yields
\begin{equation}\label{eq:finite_delta_source_coefficient_f}
\begin{split}
      f_\Delta(k)u^{1/2+i k}\equiv&\int \frac{du_0 dx_0}{u_0^2} K(0,u,x|u_0,x_0) u_0^{1/2+i k}\\
      &=\frac{2^{\Delta-3}\left|\Gamma\left(\frac{\Delta+i k}{2}-\frac{1}{4}\right)\right|^2}{\sqrt{\pi}\Gamma(\Delta-1)}u^{1/2+i k}
\end{split}
\end{equation}
Differentiating the propagator with respect to $\tau_E$ (before setting $\tau_E=0$) and repeating the same procedure as before, we find
\begin{equation}\label{eq:finite_delta_source_coefficient_g}
\begin{split}
    g_\Delta(k)u^{1/2+i k}&\equiv\int \frac{du_0 dx_0}{u_0^2} \partial_{\tau_E}K(0,u,x|u_0,x_0) u_0^{1/2+i k}\\
    &=\frac{4^{\Delta/2-1} s \left|\Gamma \left(\frac{\Delta-i k}{2}+\frac{1}{4}\right)\right|^2 }{\sqrt{\pi } \Gamma (\Delta-1)} u^{1/2+i k}
\end{split}
\end{equation}
With $u^2\pi_E=\partial_{\tau_E}\phi$ and the Lorentzian continuation $\pi_L=i\pi_E$, sources $\lambda^\pm$ on $\mathcal B_\pm$ give $\varphi_k=f_\Delta(\lambda^-_k+\lambda^+_k)$ and $\pi_k=i\,g_\Delta(\lambda^+_k-\lambda^-_k)$.
At large $\Delta$, $f_\Delta \rightarrow 1/2$ and $g_\Delta\simeq \Delta/2$, which reproduces $\varphi\simeq\frac12(\lambda^-+\lambda^+)$ and $\pi_\phi\simeq\frac{i\Delta}{2u^2}(\lambda^+-\lambda^-)$ of Section~\ref{sec3.1:scalar_field}. Since $\bar e_k=e_{-k}$, it is convenient to write $\lambda^-(u)=X(u)+iY(u)$ with $X,Y$ real. The reflection condition~\eqref{eq:reality_condtion} then gives $\lambda^\mp_k=X_k\pm iY_k$, with $X_{-k}=X_k^*$ and $Y_{-k}=Y_k^*$, so that
\begin{equation}\label{eq:finite_delta_XY}
    \varphi_k=2 f_\Delta(k)X_k,\qquad \pi_k=2g_\Delta Y_k,
\end{equation}
so the real part of the source displaces the field and the imaginary part the momentum.

Since $\int\frac{du}{u^2}\,e_k e_{k'}=\delta(k+k')$, the symplectic form~\eqref{eq:symplectic_form_scalar} pairs the mode $k$ with $-k$. With~\eqref{eq:finite_delta_XY},
\begin{equation}\label{eq:finite_delta_omega}
  \Omega_\partial(\delta_1,\delta_2)
  =\int dk\,\bigl(\delta_1\pi_k\,\delta_2\varphi_{-k}-\delta_2\pi_k\,\delta_1\varphi_{-k}\bigr)
  =4\int dk\,f_\Delta g_\Delta\,\bigl(\delta_1Y_k\,\delta_2X_{-k}-\delta_2Y_k\,\delta_1X_{-k}\bigr).
\end{equation}
The complex structure $J_{\rm FS}$ of Section~\ref{subsec:fs_boundary_symplectic} acts as $\lambda^-_k\mapsto -i\lambda^-_k$ or, in terms of the real variables, $(X_k,Y_k)\mapsto(Y_k,-X_k)$. The Fubini-Study metric $g_{\rm FS}=\frac{1}{2}\Omega_\partial(\delta_1,J_{\rm FS}\delta_2)$ and the cost metric $G_V=\frac{1}{2}\mathcal H_V$ obtained from~\eqref{eq:finite_delta_generator} are then
\begin{align}
    g_{\rm FS}(\delta_1,\delta_2)=&2\int dk\,f_\Delta g_\Delta\left(\delta_1 X_k\delta_2 X_{-k}+\delta_1 Y_k\delta_2 Y_{-k}\right),\label{eq:finite_delta_fs}\\
    G_V(\delta_1,\delta_2)=&16\pi\int dk\,f_\Delta^2\,
     \bigl(\kappa_k^2\,\delta_1X_k\,\delta_2X_{-k}+r_\Delta^2\,\delta_1Y_k\,\delta_2Y_{-k}\bigr),
  \label{eq:finite_delta_cost}
\end{align}
with   $r_\Delta(k)=\frac{g_\Delta(k)}{f_\Delta(k)}$ and $\kappa_k^2=m^2+\sigma_k$. Both metrics are diagonal in $(X_k,Y_k)$, so by~\eqref{eq:principal_penalties} the principal directions are $\delta X_k$ and $\delta Y_k$ and the principal penalties are the ratios of the diagonal entries.
\begin{equation}
    p_{X,k}=8\pi \frac{\kappa^2_k}{r_\Delta},\qquad p_{Y,k}=8\pi r_\Delta,\qquad
  r_\Delta(k)=2\left|
       \frac{\Gamma\!\left(\frac{\Delta+ik}{2}+\frac14\right)}
            {\Gamma\!\left(\frac{\Delta+ik}{2}-\frac14\right)}
     \right|^2 .
  \label{eq:finite_delta_penalties}
\end{equation}
They are positive but generally unequal. Numerically, we found that  for $\Delta\in[2,100]$ and $k\in [0,300]$, $p_{Y,k}>p_{X,k}$, supporting the fact that preparing a displacement in momentum is more expensive than a displacement in the field. Thus, a single scalar operator already produces a cost metric which is not proportional to the Fubini-Study metric.

The origin of the unequal penalties is more transparent in the coordinates $(\varphi_k,\pi_k)$. The linearised flow $A$, defined by $\Omega_\partial(\delta_1,A\delta_2)=\mathcal H_V (\delta_2,\delta_1)$, follows from the first form of~\eqref{eq:finite_delta_omega} and from~\eqref{eq:finite_delta_generator}, and $J_{\rm FS}$ follows from $(X_k,Y_k)\mapsto(Y_k,-X_k)$ and~\eqref{eq:finite_delta_XY}. Acting on $(\varphi_k,\pi_k)^T$
\begin{equation}
  A_{V,k}=8\pi
    \begin{pmatrix}0&1\\-\kappa_k^2&0\end{pmatrix},\qquad
  J_{V,k}=\begin{pmatrix}0&\kappa_k^{-1}\\-\kappa_k&0\end{pmatrix},\qquad
  J_{{\rm FS},k}=\begin{pmatrix}0&r_\Delta^{-1}\\-r_\Delta&0\end{pmatrix}.
  \label{eq:finite_delta_complex_structures}
\end{equation}
Here $A_{V,k}$ describes the compensating scalar deformation $\delta'$ written in modes. Notice that the expression for $\delta'\varphi$ and $\delta'\pi_\phi$ given above become $\delta'\varphi_k= 8\pi\pi_k$ and $\delta'\pi_k=-8\pi \kappa^2_k \varphi_k$ which are Hamilton's equations for a harmonic oscillator. As explained in Section~\ref{subsec: check_volume} $A_{V,k}$ is the linearisation of $\delta'$. Hence $A_{V,k}^2=-\omega_k^2$ with frequency $\omega_k=8\pi\kappa_k$, so that $|A_{V,k}|=\omega_k$ and $J_{V,k}=A_{V,k}|A_{V,k}|^{-1}$. The two complex structures have the same form but with $\kappa_k$ in place of $r_\Delta$, and they coincide if $\kappa_k=r_\Delta(k)$, which does not happen for any mode.

Inserting \eqref{eq:finite_delta_complex_structures} in~\eqref{eq:penalty_operator} gives $-J_{{\rm FS},k}A_{V,k}=8\pi\ \textup{diag}(\kappa_k^2/r_\Delta,r_\Delta)$, which reproduces~\eqref{eq:finite_delta_penalties}. In terms of the sources, the boundary observable reads
\begin{equation}
H_V^{(2)}=8\pi\int dk\,f_\Delta^2\Bigl[(\kappa_k^2+r_\Delta^2)\,\lambda^-_k\lambda^+_{-k}+\frac{\kappa_k^2-r_\Delta^2}{2}\bigl(\lambda^-_k\lambda^-_{-k}+\lambda^+_k\lambda^+_{-k}\bigr)\Bigr].
  \label{eq:finite_delta_same_half}
\end{equation}
The terms quadratic in sources on the same half, absent in~\eqref{eq:delta U}, vanish if $\kappa_k=r_\Delta(k)$, \textit{i.e.} if $J_V=J_{\rm FS}$ in that mode. At fixed $k$ and large $\Delta$, $r_\Delta=\nu+\sigma_k/(2\nu)+O(\nu^{-3})$ with $\nu\equiv \Delta-1$ and $p_{Y,k}/p_{X,k}=1+\nu^{-2}+O(\nu^{-4})$. Both penalties approach $8\pi \Delta$ and $J_V\to J_{\rm FS}$, recovering Section~\ref{subsec: check_volume}. The unequal weights first appear at relative order $1/\Delta^2$.

\subsection{Volume plus scalar contribution}\label{sec:dressed_volume}
In Sections \ref{subsec: check_volume} and \ref{sec:finite_dimension_cost} the preparation penalties are entirely fixed by the volume observable and the conformal dimension of the scalar $\Delta$. To show how the choice of bulk observable enters in the cost geometry, we consider a deformation of volume adding a field dependent term and follow the response of the penalties. Again, we make the same assumptions as in Section~\ref{sec:finite_dimension_cost} ($d=2$, $\ell=1$, sources independent of $x$, Ward-identity cancellation) and consider
\begin{equation}\label{eq:dressed_observable}
    W=\int_\Sigma d^d x \frac{\sqrt{h}}{G_N}+W_m
\end{equation}
with 
\begin{equation}
    W_m=\int_\Sigma d^dx\sqrt{h}\left(a_0\varphi^2+a_1 h^{ij}\partial_i\varphi\partial_j \varphi\right).
\end{equation}
where $a_0$ and $a_1$ are real constants and, as in Section~\ref{sec:finite_dimension_cost}, $\varphi$ is the restriction of $\phi$ to the surface $\Sigma$. In the notation of~\eqref{eq:complexity-anything}, this observable corresponds to $F_1=1+G_N\left(a_0\varphi^2+a_1 h^{ij}\partial_i\varphi\partial_j\varphi\right)$. Since $F_1$ also depends on the scalar field and not just on the geometry and the embedding, $W$ lies outside the purely geometric family in~\cite{Belin:2021bga,Belin:2022xmt}. Observables involving matter fields have also been considered in JT gravity~\cite{Caceres:2025myu}, more concretely a dilaton field. We are not proposing \eqref{eq:dressed_observable} as a holographic dual of complexity but we use it as a controlled deformation of volume which admits an exact treatment at quadratic order in the sources. 

The observable is only fixed once the surface $\Sigma$ is specified. This is done by specifying a concrete function $F_2$ in~\eqref{eq:complexity-anything-surface}. The two natural choices are $F_2=1$, for which $\Sigma$ is the maximal-volume slice, and $F_2=F_1$, for which $\Sigma$ extremises $W$. The argument of Section~\ref{sec3.1:scalar_field} applies to both. For $F_2=F_1$, the matter term adds to the first variation of the functional a contribution quadratic in the sources, so the extremal slice is still displaced by $O(\lambda^2)$ and the observable changes by $O(\lambda^4)$. For $F_2=1$, $W_m$ is not extremised, but both its first variation and the displacement are $O(\lambda^2)$, so the change in $W$ is again $O(\lambda^4)$. At quadratic order we may therefore take $\Sigma=\{\tau=0\}$, and we choose $F_2=F_1$ so that $\Sigma$ can be held fixed in the Peierls construction.

The deformed action $S-\epsilon W$ modifies the metric equations through the ``volume'' shell of Section~\ref{sec3.1:scalar_field}, plus an additional shell of order $\epsilon\lambda^2$ coming from the metric dependence of $W_m$. It also modifies the scalar equation of motion, which becomes
\begin{equation}\label{eq:dressed_scalar_eom}
    \sqrt{-g}\left(\Box-m^2\right)\phi=\epsilon\delta(\tau)\mathcal J,\qquad \mathcal{J}\equiv 2\sqrt{h}a_0\phi-2a_1\partial_i \left(\sqrt{h}h^{ij}\partial_j \phi\right).
\end{equation}

In WdW coordinates $\sqrt{-g}=\sqrt{h}$ and $\pi_\phi=\sqrt{h}\partial_\tau \phi$ by~\eqref{eq:conjugate_momentum} and
\begin{equation}
    \sqrt{-g}\, \Box\phi=-\partial_\tau\pi_\phi+\partial_i\left(\sqrt{h}h^{ij}\partial_j\phi\right).
\end{equation}
Let us study in more detail the variations generated by the perturbation $\epsilon W$ on both $\varphi$ and $\pi_\phi$. With this object, we write $\phi$ as a distribution
\begin{equation}
    \phi=\phi_-(\tau,\sigma)\Theta(-\tau)+\phi_+(\tau,\sigma)\Theta(\tau),
\end{equation}
while its two first derivatives read
\begin{equation}
\begin{split}
    \partial_\tau\phi&=\{\partial_\tau \phi\}+[\phi]\delta(\tau)\\
    \partial_\tau^2\phi&=\{\partial_\tau^2\phi\}+[\partial_\tau\phi]\delta(\tau)+[\phi]\delta'(\tau),
    \end{split}
\end{equation}
where $\{A\}$ denotes the regular part of $A$ and $[A]$ its discontinuity. Since the right-hand side of~\eqref{eq:dressed_scalar_eom} contains $\delta(\tau)$ but not $\delta'(\tau)$ (as it appears on the left-hand side), $[\phi]=0$ and thus $\phi$ is continuous along $\tau=0$. Integrating~\eqref{eq:dressed_scalar_eom} over $\tau\in(-\eta,\eta)$ and taking $\eta\rightarrow0$ gives the following jump on the conjugate momentum
\begin{equation}
    [\pi_\phi]=-\epsilon \mathcal J,
\end{equation}
Writing $\phi_{R,A}=\phi_0+\epsilon\delta_{R,A}\phi$, the retarded perturbation vanishes for $\tau<0$ and has $(\delta_R\varphi,\delta_R\pi_\phi)=(0,-\mathcal J)$ at $\tau=0^+$, while the advanced solution vanishes for $\tau>0$ and has $(\delta_A\varphi,\delta_A\pi_\phi)=(0,\mathcal J)$ at $\tau=0^-$. Their difference~\eqref{eq:peierls_variation_summary} is continuous across $\Sigma$ and defines a homogeneous solution with
\begin{equation}\label{eq:dressed_matter_flow}
    \delta_m\varphi=0,\qquad\delta_m\pi_\phi=-\mathcal J,\qquad \text{on }\Sigma.
\end{equation}
With the scalar symplectic form \eqref{eq:symplectic_form_scalar}
\begin{equation}\label{eq:dressed_matter_pairing}
    \Omega_{\rm scalar}(\delta,\delta_m)=\int_\Sigma du\, dx(\delta\pi_\phi\delta_m\varphi-\delta_m\pi_\phi\delta\varphi)=\int_\Sigma \mathcal J\delta\varphi=\delta W_m
\end{equation}
up to terms in which $\delta$ acts on the metric inside $W_m$. These terms are of order $\lambda^2\delta h=O(\lambda^3\delta\lambda)$ and can be ignored.

The remaining pieces of the flow are unchanged at this order. The matter shell modifies the jump of the extrinsic curvature but at order $O(\lambda^2)$ and leaves $h_{ij}$ continuous. In this pairing, it multiplies $\delta h_{ij}\delta\pi^{ij}$ and thus contributes order $\lambda^3\delta\lambda$. The total flow is again given by the York deformation together with $\delta'$ of Section~\ref{sec:finite_dimension_cost} plus $\delta_m$. The diffeomorphism part cancels in the regulated pairing, as happened before. 

In the large-$\Delta$ approximation of Section \ref{sec3.1:scalar_field}, the scalar initial data are $\varphi\simeq \frac{1}{2}(\lambda^++\lambda^-)$ and $\pi_\phi\simeq \frac{i\Delta}{2u^2}(\lambda^+-\lambda^-)$. Writing $\lambda^-=X+iY$ and imposing the reflection condition $\lambda^+= (\lambda^-)^*$ they become
\begin{equation}\label{eq:dressed_X_Y}
    \varphi= X,\qquad \pi_\phi=\frac{\Delta}{u^2} Y
\end{equation}
The real part of the source thus prepares a displacement of the field, while the imaginary part prepares a displacement of the momentum. In these variables, the boundary symplectic form reads
\begin{equation}
    \Omega_\partial(\delta_1,\delta_2)=\Delta\int \frac{du\,dx}{u^2}\left(\delta_1 Y\delta_2 X-\delta_2 Y\delta_1 X\right).
\end{equation}
In $d=2$ one has $\sqrt{h}h^{ij}=\delta^{ij}$, so for $x-$independent sources, the gradient term of $W_m$ reduces to $\int du(\partial_u\varphi)^2=-\int \frac{du}{u^2} u^2 \varphi \partial_u^2\varphi$ after integration by parts with vanishing boundary terms. Together with the volume piece, the generator becomes
\begin{equation}\label{eq:dressed_generator}
    H_W^{(2)}=4\pi \Delta^2\int \frac{du\,dx}{u^2}\left(X^2+Y^2+X\hat{\mathcal D}  X\right)
\end{equation}
with $\hat{\mathcal{D}}\equiv (a_0-a_1 u^2\partial_u^2)/(4\pi \Delta^2)$. By~\eqref{eq:dressed_matter_flow}, the matter part of the transported flow only shifts the momentum. In terms of the sources, it reads
\begin{equation}
    \delta_m\lambda^\pm=\pm\frac{i}{\Delta}(a_0-a_1u^2\partial_u^2)(\lambda^++\lambda^-)
\end{equation}
to which we have to add the phase rotation $\delta'\lambda^\pm=\pm 8\pi i \Delta\lambda^\pm$ of Section~\ref{subsec: check_volume}. In contrast to the latter, this one mixes both $\lambda^+$ and $\lambda^-$ but still preserves the reflection condition because $X$ is real.

Expanding $\lambda^\pm$ in the basis~\eqref{eq:basis}, every operator appearing in this example becomes diagonal and $\mathcal{D}$ and $\hat{\mathcal{D}}$ are respectively replaced by 
\begin{equation}
    w_k=a_0+a_1\sigma_k,\qquad \hat w_k=\frac{a_0+a_1\sigma_k}{4\pi \Delta^2}.
\end{equation}
The reference state $\lambda^\pm=0$ is a critical point, since \eqref{eq:dressed_generator} is quadratic. The Hessian is positive and bounded below if $4\pi \Delta^2(1+\hat w_k)$ is positive and bounded  away from zero for all $k$, which happens if $a_1\geq 0$ and $4\pi \Delta^2+a_0+a_1/4>0$. If any of these two assumptions fail, the observable does not define a positive cost metric around the vacuum.

For each mode, in the basis $(X_k,Y_k)$ , the map $A$, its modulus and the associated complex structure are
\begin{equation}\label{eq:dressed_flow_J}
	\begin{split}
		    A_{W,k}=8\pi \Delta\begin{pmatrix}
			0 & 1\\
			-(1+\hat w_k)& 0
		\end{pmatrix},\qquad |A_{W,k}|=8\pi \Delta&\sqrt{1+\hat w_k},\\
		J_{W,k}=\begin{pmatrix}
			0 & (1+\hat w_k)^{-1/2}\\
			-(1+\hat w_k)^{1/2}&0
		\end{pmatrix}&
	\end{split}
\end{equation}
It is immediately checked that $\Omega_\partial(u,A_Wv)=\mathcal{H}(v,u)$, that $J_{W,k}^2=-1$ and that $\Omega_\partial(J_Wu,J_W v)=\Omega_\partial(u,v)$ as required by the construction in Section~\ref{sec4.1: observable_metric}. For $\hat w_k=0$, $J_{W,k}$ acts as $(X,Y)\rightarrow (Y,-X)$ or likewise $\lambda^\pm\rightarrow \pm i\lambda^\pm$ of $J_V$. The cost metric is
\begin{equation}\label{eq:dressed_cost}
    \begin{split}
    G_W(\delta_1,\delta_2)=&4\pi \Delta^2\int \frac{du\,dx}{u^2}\left(\delta_1 X(1+\hat{\mathcal D})\delta_2 X+\delta_1 Y\delta_2 Y\right)\\
   =&2\pi \Delta^2\int\frac{du\,dx}{u^2}\Bigl[\delta_1\lambda^+\delta_2\lambda^-+\delta_2\lambda^+\delta_1\lambda^-
    +\tfrac12\bigl(\delta_1\lambda^++\delta_1\lambda^-\bigr)\,\hat{\mathcal D}\,\bigl(\delta_2\lambda^++\delta_2\lambda^-\bigr)\Bigr],
    \end{split}
\end{equation}
which is symmetric since $\hat{\mathcal D}$ is self-adjoint. The first two terms in the second line are the volume cost metric of Section~\ref{subsec: check_volume} and the last one is the contribution of the scalar piece. The Kähler metric of $J_W$ is
\begin{equation}
    g_J(\delta_1,\delta_2)=\Delta\int\frac{du\,dx}{u^2}\Bigl[\delta_1X\,(1+\hat{\mathcal D})^{1/2}\,\delta_2X+\delta_1Y\,(1+\hat{\mathcal D})^{-1/2}\,\delta_2Y\Bigr],
\end{equation}
where $(1+\hat{\mathcal{D}})$ acts mode by mode as a multiplication by $(1+\hat w_k)^{\pm1/2}$. For $a_0=a_1=0$, we recover the cost metric in Section~\ref{subsec: check_volume}. For each mode, $G_W(\delta_1,\delta_2)=4\pi \Delta\sqrt{1+\hat{w}_k}g_J(\delta_1,\delta_2)$, so the cost metric is still proportional to the Kähler metric of $J_W$ within each mode. If $a_1\neq 0$, however, the frequencies depend on $k$, so this proportionality does not hold globally.

We now compute the penalties defined in Section~\ref{sec4.1: observable_metric}. At the order we work, $J_{\rm FS}=J_V$, so they are the eigenvalues of $-J_V A_W$. Using~\eqref{eq:dressed_flow_J} and the fact that in the coordinates $(X_k,Y_k)$, $J_V$ acts as 
$$J_V=\begin{pmatrix}0& 1\\ -1&0\end{pmatrix}$$
one finds
\begin{equation}
    -J_V A_{W,k}=8\pi \Delta \begin{pmatrix}
        1+\hat{w}_k & 0\\
        0 & 1
    \end{pmatrix}
\end{equation}
which is diagonal. Equivalently, $G_W$ in \eqref{eq:dressed_cost} and the Fubini-Study metric are simultaneously diagonal in the same basis. The penalised directions are therefore variations of the real and imaginary part of the sources with penalisations
\begin{equation}
    \begin{split}
        p_{X,k}=8\pi \Delta\left(1+\hat{w}_k\right),\quad p_{Y,k}=8\pi \Delta.
    \end{split}
\end{equation}
We emphasise several features.  First, the dressing only weights variations of the real part of the source, which by~\eqref{eq:dressed_X_Y} displace the field. This is a consequence of the fact that $W_m$ depends on $\varphi$ but not on $\pi_\phi$ so the flow generated by this observable only shifts the momentum. Variations of the imaginary part, which involve the momentum, keep the penalty fixed.

Second, the two couplings act differently across modes. The coupling $a_0$ shifts $p_{X,k}$ uniformly in $k$. In contrast, $a_1$ introduces a dependence on the mode $k$ in such a way that modes with large $k$ are more penalised than those with small $k$ and therefore are more expensive to turn on. 

\section{Discussion}\label{sec:discussion}

In the present manuscript we have developed a local prescription that associates suitable bulk observables with quadratic geometries of boundary state preparation. Our starting point is the established holographic transport of Hamiltonian flows, which uses the equality of bulk and boundary symplectic forms to associate a boundary generator $H_W^{\rm CFT}$ with a bulk observable $W_{\rm bulk}$ \cite{Belin:2018bpg,Belin:2022xmt}. We have used this construction to address the additional question of how the choice of observable determines preparation costs. The principal results are an explicit calculation of finite conformal-dimension penalties for the maximal volume and a controlled example showing how scalar and gradient additions to volume modify those penalties.

The cost prescription applies near a reference state $\lambda_i$ at which $H_W^{\rm CFT}$ is stationary and has a positive-definite Hessian $\mathcal{H}$. Taking $G = \frac{1}{2}\mathcal{H}$ defines a positive quadratic cost on the allowed source variations. With the metric held fixed in source coordinates and the normalisation adopted in Section \ref{sec4: flow_to_cost}, the minimising preparation path is a straight line traversed at constant speed and its energy satisfies
\begin{equation}
    \mathcal C(\lambda_i,\lambda_f)
    =H_W^{\rm CFT}(\lambda_f)-H_W^{\rm CFT}(\lambda_i)+O(v^3),
    \qquad v=\lambda_f-\lambda_i.
\end{equation}
Consequently, $\delta \mathcal C = \delta W_{\rm bulk}$ holds to leading order around the reference state. This agreement follows by construction. The physical content lies in the preparation penalties selected by the observable, which we compare using the Fubini-Study metric. Directions with equal Fubini-Study norm produce the same leading change in fidelity, so unequal principal penalties distinguish preparation costs for variations with equal state distinguishability.

For the maximal-volume example, we first recover the large-conformal dimension scalar result of \cite{Belin:2018bpg}, for which the cost metric is proportional to the Fubini-Study metric and all directions in the scalar sector carry the same relative penalty. We then retain the full conformal-dimension dependence at quadratic order in the $x$-independent scalar sector around the AdS$_3$ vacuum. For $\Delta \geq 2$, the resulting principal penalties are positive but generally unequal for variations of the real and imaginary parts of the source. These variations displace the bulk scalar field and its conjugate momentum, respectively, giving a direct interpretation of the two preparation costs. The volume cost metric therefore does not need to be proportional to the Fubini-Study metric even within the single scalar sector. Correspondingly, the complex structure $J_V$ selected by the volume generally differs from $J_{\rm FS}$. At fixed mode, the relative splitting of the penalties first appears at order $1/\Delta^2$, while the common penalty and the equality of complex structures are recovered at leading order in the large-dimension limit.

The second calculation demonstrates how changing the observable changes the preparation geometry. We deform volume by quadratic scalar field and spatial gradient terms and compute their effects at quadratic order in the sources, using the large-conformal dimension approximation for the source-to-initial-data map. Both additions modify the field displacement penalties while the momentum displacement penalties remain unchanged at the order considered. The scalar field term shifts the field penalties equally across modes, whereas the gradient term introduces mode dependence. We also determine the coupling restrictions required for the resulting quadratic form to remain positive. This example makes the observable dependence of the penalties explicit: the state-space geometry used to compare directions is fixed, while changing the bulk functional changes their relative preparation costs. The matter-dependent functional extends beyond the purely geometric family considered in \cite{Belin:2022xmt} and serves here as a controlled test of the cost prescription, establishing it as a holographic complexity dual would require further evidence.

The transport calculations provide supporting checks of the framework. For maximal volume, the Peierls construction reproduces the York deformation and its boundary representation, in agreement with \cite{Belin:2018bpg,Belin:2022xmt}. In Appendix \ref{app:B}, the flow generated by $\mathcal A/4G_N$ for the extremal surface associated with a vacuum half-line is represented by a one-sided boost \cite{Kaplan:2022orm} (see also \cite{Dong:2025orj} for the Peierls construction of these flows for geometric entropies). Its boundary symplectic pairing yields $\delta \mathcal A/4G_N = \delta \braket{K_{\mathcal A}}$ and hence recovers the vacuum entanglement first law through the RT relation. This example checks the identification of the transported boundary functional without requiring the additional assumptions needed for a positive preparation cost.

The meaning of the boundary generators remains an important distinction. Symplectic transport determines $\delta H_W^{\rm CFT} = \delta W_{\rm bulk}$ and hence fixes $H_W^{\rm CFT}$ locally up to an additive constant. The resulting quantity is a function on the chosen family of source prepared states. It does not necessarily need to be the expectation value of a state independent CFT operator, as the entanglement entropy example illustrates. Its first variation can nevertheless admit an operator interpretation at a fixed reference state: for entropy, this is supplied by the modular Hamiltonian of that state. Identifying similarly meaningful CFT operators for the first variations of more general transported observables remains an open direction.

The Hamiltonian flow generated by $H_W^{\rm CFT}$ also has a different role from the optimal preparation path. It preserves the generator, whereas preparation generally connects states with different values of that generator. In the quadratic approximation, the linearised flow $A$ determines a compatible complex structure $J_W$ and the terminal velocity of the optimal path is related to the transported flow by
\begin{equation}
    (s_f-s_i)\dot\lambda_*(s_f)
    =-\lvert A\rvert^{-1}J_W\delta_W^\partial\big|_{\lambda_f} +O(v^2).
\end{equation}
The rotation by $J_W$ and rescaling by the inverse frequencies yield the radial displacement from the reference state to leading order. This relation explains how the transported flow determines the directions of the optimal source paths in the frozen quadratic metric. The finite-dimension and deformed volume examples show why $J_W$ must be distinguished from the complex structure supplied by the Fubini-Study geometry.

Our prescription provides a way to organise the freedom in choosing bulk observables through their associated preparation penalties. Its applicability depends on the reference state and the allowed family of variations: stationarity and Hessian positivity are essential for the positive quadratic cost, although they are not required for symplectic transport itself. When the Hessian has null directions, redundancies in the source description must be distinguished from genuine degeneracies of the observable. An indefinite Hessian cannot directly define the cost considered here. The resulting association is also not one-to-one. Distinct bulk observables whose boundary generators have the same Hessian at the reference state yield the same quadratic cost geometry, even when the higher-order dependence differs. The coupling restrictions in the deformed volume example give an explicit illustration of how admissibility constrains the choice of observable.

Several limitations remain. The construction fixes the metric at the reference state and extending it with constant components in source coordinates suffices for the quadratic calculation. This extension is not uniquely selected by transport: other smooth metrics agreeing at the reference state give the same leading preparation cost but may change higher order terms and optimal paths. Matching the optimised cost to the boundary generator beyond the quadratic order requires control of operator domains, UV divergences, contact terms and boundary contributions. In our examples, decay assumptions on the sources and a regulated treatment of the Ward identities are used to define the relevant pairings. The York deformation makes this particularly relevant near the preparation slice, where its boundary Weyl rescaling is discontinuous. Establishing which features of the penalties are independent of the regulator and renormalisation scheme is an important further test.

The new calculations suggest several concrete extensions. The finite-dimensional analysis can be generalised to spatially dependent sources and the scalar deformation of volume can be studied with the full finite-dimension source-to-initial-data map. This would allow the two effects isolated here to be compared within the same calculation. Other bulk observables and reference backgrounds would test how broadly positive cost geometries arise. In particular, extending the analysis to two-sided black holes could clarify the relation to interior growth, provided the stationarity and positivity assumptions are checked for the chosen reference state and preparation family. Higher derivative and quantum corrections offer further tests of both transport and the resulting penalties. A separate physical task is to identify classes of boundary preparation operations whose costs are described by these source-space metrics. Such an identification is needed to connect the local preparation energy to an operational notion of circuit complexity.

The finite-dimension volume calculation and the scalar deformation show explicitly how a bulk observable can distinguish the costs of boundary variations with equal state distinguishability. They supply a local relation between the choice of observable and the assignment of preparation penalties, with positivity providing a concrete admissibility criterion. Extending this relation to physically specified preparation operations and beyond the quadratic neighbourhood of the reference state is the next step towards a broader complexity interpretation.

\section*{Acknowledgments}
We are grateful to J.L. Barbón, E. Cáceres, A. Karch, D. Neuenfeld, J. Pedraza, S.M. Ruan, A. Speranza and A. Svesko for many enlightening discussions. We especially thank E. Cáceres, A. Karch and A. Svesko for their comments on the manuscript. This publication has been funded within the framework of the R\&D\&I Projects PID2024-156043NB-I00 and CEX2025-001574-S, funded by MICIU/AEI/10.13039/501100011033. The research presented in this publication falls within the research line Strings and Quantum Gravity.
RC is supported by the Spanish grant FPU with reference FPU22/01262 and also acknowledges the additional support from the CSIC grant IMOVE24101 and the hospitality of the University of Texas at Austin during the beginning of this project.
The work of CP was supported by the U.S. Department of Energy under grant number DESC0019470 and by the Heising-Simons Foundation under the ``Observational Signatures
of Quantum Gravity” collaboration grants 2021-2818 and 2024-5305.

\appendix
\section{Review of the covariant phase space formalism} \label{app1:Review}

The covariant phase space formalism provides a Hamiltonian description of a field theory preserving the covariance of spacetime \cite{Crnkovic:1986ex, Lee:1990nz, Iyer:1994ys}. Instead of splitting spacetime into space and time and working with initial data localised on a preferred time slice, one takes the phase space to be the set of solutions that satisfy the equations of motion while the symplectic structure is directly built from the Lagrangian. This formalism has been extensively employed in the context of gravity, where there is no preferred notion of time and all the observables are diffeomorphism invariant. Over the last decades, it has become the standard approach to determine gravitational charges \cite{Ashtekar:1981bq, Compere:2017knf, Wald:1999wa,Barnich:2001jy} and to derive the so-called ``first law of black hole mechanics'' \cite{Wald:1993nt,Iyer:1994ys}, and it underlies the derivation of linearised Einstein's equations from the ``first law of entanglement'' \cite{Faulkner:2013ica}.

There are two main reasons why this formalism plays a central role here. First, it allows treating any differentiable bulk observable as a Hamiltonian and associating with it a flow in the space of solutions which, in general, has nothing to do with time evolution. Second, as we explain in Section \ref{sec2:holographic_transport}, the symplectic form obtained in this way for the bulk theory $\Omega$ can be related to a different symplectic form $\Omega_\partial$ constructed solely from CFT data. This identification is what will allow us to determine the boundary generator $H_W$ dual to the bulk observable. In this appendix we will review the main ingredients that are needed for addressing these two points and fix the conventions used throughout the paper. This review is not intended to be exhaustive. We refer the interested reader to \cite{Lee:1990nz,Harlow:2019yfa, Compere:2019qed} for further details.  

A symplectic manifold $(\mathcal{P},\Omega)$ is a manifold endowed with a two-form $\Omega$ that is (\textit{i}) closed ($\delta\Omega=0$) and (\textit{ii}) non-degenerate \textit{i.e.} $\Omega(\delta_1,\delta_2)=0$ for all $\delta_2$ implies $\delta_1=0$. This $\Omega$ is called symplectic form. Henceforth, $\delta$ (without any labels) represents an exterior derivative in this manifold, so that closedness reads $\delta\Omega=0$, while labelled symbols as $\delta_{1,2}$ denote a tangent vector of $\mathcal{P}$. In finite dimensions, every smooth function $H:\mathcal{P}\to\mathbb{R}$ determines a unique Hamiltonian vector field. In field theory the symplectic form is, in general, only weakly non-degenerate (the map from tangent to cotangent vectors is injective but not surjective) \cite{MarsdenRatiu:1999,Chernoff:1974}, so we restrict to observables for which such a vector field exists. Non-degeneracy of the symplectic form then guarantees its uniqueness. We denote this vector field by $\delta_H$, with defining relation
\begin{equation}\label{eq: hamilton_eq}
    \delta H=\Omega(\cdot,\delta_H),\quad \text{i.e.}\quad \delta_1H=\Omega(\delta_1,\delta_H),\quad \forall\delta_1.
\end{equation}
The integral lines of this vector field $\delta_H$ describe a trajectory in $\mathcal{P}$ in the same way as Hamilton's equations give the evolution of a system in phase space. For this reason, we can understand this $H$ as a Hamiltonian and $\delta_H$ as a Hamiltonian vector field. We stress that $H$ can be any function, not only an energy functional and, thus the trajectory it defines in phase space in general does not correspond to time evolution. 

Likewise, the Poisson bracket admits a natural representation in terms of the symplectic form. Indeed, given two functions $f,g:\mathcal{P}\rightarrow \mathbb{R}$, with Hamiltonian vector fields $\delta_f,\delta_g$ respectively
\begin{equation}
    \{f,g\}=\Omega(\delta_g,\delta_f),
\end{equation}
so that $\dot f=\{f,H\}$ along the flow of $\delta_H$ and for $\Omega=\delta p\wedge \delta q$, $\{q,p\}=1$.

The next question is how this structure can be defined for a given field theory. Let this theory be characterised by some Lagrangian $L$, which we consider to be a $(d+1)$-form in a Lorentzian spacetime $\mathcal{M}$, depending on a set of fields $\Phi$ and their derivatives. Defining a symplectic manifold associated with the given theory is somewhat subtle for reasons that will be explained momentarily. For the time being, we define the \textit{pre-phase space} $\tilde{\mathcal{P}}$ as the set of  solutions to the equations of motion. A tangent vector $\delta_1$ at a solution $\Phi_0$ is characterised by its components $\delta_1 \Phi \equiv \delta\Phi(\delta_1)$, \textit{i.e.} by the infinitesimal change of the fields along that vector. 

In order to construct a symplectic form, we start from the variation of the Lagrangian. After the application of the chain rule as many times as needed, it reads
\begin{equation} \label{eq:bulk_action_vars}
    \delta L=E_{\Phi}(\Phi)\delta\Phi+d\theta(\Phi,\delta\Phi),
\end{equation}
where $E_\Phi(\Phi)$ are the equations of motion and $\theta$ is the so-called symplectic potential. The latter is a $d$-form in spacetime and a 1-form in pre-phase space, and it is constructed out of the fields, their derivatives and their variations but does not explicitly depend on the coordinates. The variation of this symplectic potential defines the so-called pre-symplectic current (for a detailed treatment of boundaries in this formalism we refer to \cite{Harlow:2019yfa})
\begin{equation}
    \tilde{\omega}=\delta \theta,\qquad \textup{i.e.}\qquad \tilde{\omega}(\delta_1,\delta_2)=\delta_1\theta(\delta_2)-\delta_2\theta(\delta_1),
\end{equation}
for commuting $\delta_1$ and $\delta_2$. This is a $d$-form in spacetime and a two-form in $\mathcal{P}$. An important property that will be used repeatedly is that $\tilde\omega$ is closed for both the exterior derivative in spacetime $d$ (on-shell) and the exterior derivative in pre-phase space $\delta$. Indeed, using $\delta^2=0$ and the fact that $d$ commutes with $\delta$ 
\begin{equation}
    d\tilde\omega=d\delta \theta=\delta d\theta=\delta(\delta L-E_\Phi \delta\Phi)= - \delta E_\Phi\wedge \delta \Phi=0,
\end{equation}
where in the last step, we have considered on-shell variations of the equations of motion so $\delta E_\Phi=0$. Similarly, $\delta\tilde\omega=\delta^2 \theta=0$.

Finally, given any Cauchy surface $\Sigma$ the presymplectic form is defined as 
\begin{equation} \label{eq:symp_form}
    \tilde{\Omega}(\delta_1,\delta_2)=\int_\Sigma \tilde{\omega}(\delta_1,\delta_2).
\end{equation}
The independence of $\tilde{\Omega}$ on the Cauchy surface relies on the behaviour of the symplectic current at the asymptotic boundary. Let $\Sigma_1$ and $\Sigma_2$ be two different Cauchy surfaces, with $\Sigma_2$ in the future of $\Sigma_1$, and $\Gamma$ the region of the asymptotic (timelike) boundary between $\partial \Sigma_1$ and $\partial\Sigma_2$. Let also $\mathcal{V}$ be the volume contained between them. Stokes' theorem applied here gives
\begin{equation}
    \int_{\mathcal{V}}d\tilde{\omega}=\int_{\Sigma_2}\tilde{\omega}-\int_{\Sigma_1}\tilde{\omega}+\int_{\Gamma}\tilde{\omega},
\end{equation}
where the surfaces are oriented with future-directed normals and $\Gamma$ with outward normal. Due to the closedness of the symplectic current, the left-hand side of the previous expression vanishes. Thus, the presymplectic form defined above is conserved when the flux through $\Gamma$ vanishes. If the boundary
current is instead a total derivative, a corresponding corner contribution to $\partial\Sigma_{1,2}$ must be included in the conserved presymplectic form \cite{Harlow:2019yfa}.

Until now, we have been agnostic about the non-degeneracy of the symplectic form. For instance, in gauge theories, there exist equivalent solutions that are related by a gauge transformation. In other words, these gauge transformations are tangent vectors $\delta_X$ such that
\begin{equation}
    \tilde{\Omega}(\delta_X,\cdot\;)=0.
\end{equation}
When the former equality holds, it is said that $\delta_X$ is a zero mode. Together with the commutator, this set of transformations forms a Lie algebra. Indeed, given two zero modes $\delta_X,\delta_Y$, since $[\delta_X,\delta_Y]=\mathcal{L}_{\delta_X} \delta_Y$ one has
\begin{equation}
    \begin{split}
    \tilde{\Omega}([\delta_X,\delta_Y],\cdot\;)=&\tilde{\Omega}(\mathcal{L}_{\delta_X}\delta_Y,\cdot\;)=\mathcal{L}_{\delta_X}(\tilde \Omega(\delta_Y,\cdot\;))-\iota_{\delta_Y}\left(\mathcal{L}_{\delta_X}\tilde{\Omega}\right)\\
     &=-\iota_{\delta_Y}\left(
    \iota_{\delta_X}\delta\tilde{\Omega}
    +\delta\bigl(\tilde{\Omega}(\delta_X,\cdot)\bigr)
    \right)=0.
    \end{split}
\end{equation}
Let us denote the group of zero modes of $\tilde{\Omega}$ by $\mathcal G$. Thus, the actual phase space is defined as the quotient between the pre-phase space and this group $\mathcal G$, \textit{i.e.} $\mathcal{P}\equiv\tilde{\mathcal{P}}/\mathcal G$.

Finally, a non-degenerate definition of the symplectic form is still missing. To fix this gap, we introduce a map $\mathfrak{p}:\tilde{\mathcal{P}}\rightarrow\mathcal{P}$ that projects one solution to its $\mathcal G$-orbit. Let $\tilde{\delta}_X,\tilde{\delta}_Y$ be two tangent vectors in $\tilde{\mathcal{P}}$ and let $\delta_X,\delta_Y$ be their images under $\mathfrak{p}$. Then, we define the symplectic form $\Omega$ as
\begin{equation}
    \Omega(\delta_X,\delta_Y)\equiv \tilde{\Omega}\left(\tilde{\delta}_X,\tilde{\delta}_Y\right),
\end{equation}
which is well-defined, since $\tilde\Omega$ is invariant under the addition of zero modes and under their flow, and non-degenerate by construction. 

Two consequences of this construction will be used below. First, only gauge-invariant functions on $\tilde{\mathcal{P}}$ admit a Hamiltonian vector field since given a function $f$ and a zero mode $\delta_X$, $\delta_X f=\tilde\Omega(\delta_X,\delta_f)$. As $\delta_X$ is a zero mode, the last expression vanishes, which implies that the flow along $\delta_X$ of $f$ also does. This only happens if $f$ is gauge invariant. These Hamiltonian vector fields are defined up to the addition of zero modes. This is important because the flows generated by diffeomorphism invariant observables, such as the extremal area or maximal volume, may be supplemented by diffeomorphisms that vanish sufficiently fast at the asymptotic boundary. Second, given any diffeomorphism $\delta_\xi \Phi=\mathcal{L}_\xi \Phi$, the symplectic current $\omega(\delta,\delta_\xi)$ reduces to a total derivative on-shell \cite{Iyer:1994ys}. Thus, $\Omega(\delta,\delta_\xi)$ can be written as an integral over the boundary of the chosen Cauchy slice. Those diffeomorphisms that vanish at the boundary of $\Sigma$ are zero modes, whereas those that do not, are physical rather than a pure gauge.

To make these concepts concrete, we will sketch the calculation of the symplectic form for a scalar field and general relativity used in Section~\ref{sec3: CV} and Appendix~\ref{app:B}, where we find the boundary duals of the volume of the maximal slice and of the area of an extremal surface.

\subsection{Scalar field theories}
Before moving on to the calculation of the symplectic form of these theories, we will dedicate a few words to describe the ADM decomposition of the $(d+1)$-dimensional spacetime $\mathcal{M}$ we will work with from now on. This decomposition will play a central role since it allows us to reduce the symplectic forms that we are going to compute to a very suitable expression. 

Let us consider a global time function $t(x)$, with $x$ the coordinates in $\mathcal{M}$, whose level sets $\Sigma_t$ foliate the whole spacetime. We will consider that these surfaces have unit future-directed normal vector that we will call $n^\mu$. Moreover, the bulk metric is going to be $g_{\mu\nu}$ while the induced metric on $\Sigma_t$ will be $h_{ab}$. In ADM variables, the metric $g_{\mu\nu}$ is decomposed as \cite{Arnowitt:1959ah,Carrasco:2023fcj}
\begin{equation}
    ds^2=g_{\mu\nu}dx^\mu dx^\nu= -N^2 dt^2+h_{ab}(N^adt+dy^a)(N^bdt+dy^b).
\end{equation}
with $N$ the lapse, $N^a$ the shift and $y^a$ are the coordinates on $\Sigma_t$. 

Let us consider the action of the scalar field on a fixed background metric
\begin{equation}
\begin{split}
    S&=\int_\mathcal{M} d^{d+1}x\sqrt{-g}\left(-\frac{1}{2}g^{\mu\nu}\nabla_\mu\phi\nabla_\nu\phi-V(\phi)\right)\\
    &=\int_\mathcal{M} d^{d+1}x N\sqrt{h}\left(-\frac{1}{2}g^{tt}(\nabla_t\phi)^2-g^{ta}\nabla_t\phi\nabla_a\phi-\frac{1}{2}g^{ab}\nabla_a\phi\nabla_b\phi- V(\phi)\right)
\end{split}
\end{equation}
From this expression we can read the Lagrangian of the theory
\begin{equation}
    \mathcal{L}= N\sqrt{h}\left(-\frac{1}{2}g^{tt}(\nabla_t\phi)^2-g^{ta}\nabla_t\phi\nabla_a\phi-\frac{1}{2}g^{ab}\nabla_a\phi\nabla_b\phi- V(\phi)\right),
\end{equation}
The momentum conjugate to the scalar field is 
\begin{equation}\label{eq:conjugate_momentum}
    \pi_\phi=\frac{\partial \mathcal{L}}{\partial \dot {\phi}}=\frac{\sqrt{h}}{N}\left(\dot{\phi}-N^a\nabla_a\phi\right).
\end{equation}
The variation of this Lagrangian with respect to the scalar field yields
\begin{equation}
    \delta \mathcal{L}=\sqrt{-g}\left[(\Box \phi-V'(\phi))\delta\phi-\nabla_\mu\left(\nabla^\mu\phi\delta\phi\right)\right].
\end{equation}
From the first term on the right-hand side we recognise the equations of motion of this field while the second yields the symplectic potential
\begin{equation}
    \theta^\mu=-\sqrt{-g}g^{\mu\nu}\nabla_\nu\phi\delta\phi.
\end{equation}
The pre-symplectic current is readily obtained from the last expression after taking the antisymmetrised variation of the previous potential
\begin{equation}
    \tilde{\omega}^\mu(\delta_1,\delta_2)=-\sqrt{-g}g^{\mu\nu}(\nabla_\nu\delta_1\phi\delta_2\phi-\nabla_\nu\delta_2\phi\delta_1\phi).
\end{equation}
Recall that the pre-symplectic form requires a choice of the Cauchy slice. Here, we will choose it to be one of the $\Sigma_t$, which corresponds to the hypersurface with $t=$constant. Thus,
\begin{equation}
    \tilde{\Omega}(\delta_1,\delta_2)=\int_{\Sigma_t}\tilde{\omega}(\delta_1,\delta_2)|_{\Sigma_t}=\int_{\Sigma_t}d^dy\frac{\sqrt{h}}{N}[(\delta_1\dot\phi-N^a\delta_1\partial_a\phi)\delta_2\phi-(1\leftrightarrow2)].
\end{equation}
And from the definition of the conjugate momentum \eqref{eq:conjugate_momentum}, this expression reduces to 
\begin{equation}\label{eq:symplectic_form_scalar}
    \tilde{\Omega}(\delta_1,\delta_2)=\int_{\Sigma_t}d^dy(\delta_1\pi_\phi\delta_2\varphi-\delta_2 \pi_\phi \delta_1 \varphi),
\end{equation}
with $\varphi=\phi|_{\Sigma_t}$.

\subsection{General relativity} 
The action for general relativity supplemented with the Gibbons-Hawking-York term is
\begin{equation}
    S=\frac{1}{16\pi G_N}\int_\mathcal{M} d^{d+1}x\sqrt{-g}(R-2\Lambda)+\frac{1}{8\pi G_N}\int_{\partial\mathcal{M}} d^dx\sqrt{h} \tilde K,
\end{equation}
where $\mathcal{M}$ is the $(d+1)$-dimensional spacetime, $R$ its Ricci scalar, $\Lambda$ the cosmological constant and $\tilde K$ the trace of the extrinsic curvature tensor of the boundary $\partial \mathcal{M}$. For this computation, we again resort to the ADM decomposition of the spacetime and assume a foliation with the surfaces of constant coordinate $t$, $\Sigma_t$. To write the action in a more convenient form, we first note the Ricci scalar can be expressed in terms of the intrinsic and extrinsic curvature of $\Sigma_t$
\begin{equation}
    R= \overline{R}-(K^2-K_{ab}K^{ab})-2\nabla_\alpha (n^\beta \nabla_\beta n^\alpha- n^\alpha K).
\end{equation}
Here, $\overline{R}$ represents the Ricci scalar of the $\Sigma_t$ surface and $n^\alpha$ is the vector normal to the surface. Combining the total derivative with the Gibbons-Hawking-York term, the bulk part of the ADM action becomes \begin{equation}
    S=\frac{1}{16\pi G_N}\int_\mathcal{M}d^{d+1}x N \sqrt{h}\left(\overline{R}-K^2+K_{ab}K^{ab}-2\Lambda\right).
\end{equation}
The extrinsic curvature tensor can be expressed in terms of the induced metric $h_{ab}$, lapse and shift
\begin{equation}
    K_{ab}=\frac{1}{2N}\left(\dot{h}_{ab}-D_a N_b-D_bN_a\right),
\end{equation}
where $D_a$ is the covariant derivative associated with $h_{ab}$. The momentum conjugate to $h_{ab}$ is obtained in the same way as in the scalar case 
\begin{equation}
    \pi^{ab}=\frac{\partial \mathcal{L}}{\partial \dot h_{ab}}=\frac{\sqrt{h}}{16\pi G_N}\left(K^{ab}-K h^{ab}\right).
\end{equation}
Using the canonical ADM representative of the symplectic potential, its pullback to $\Sigma_t$ reads
\begin{equation}
    \theta|_{\Sigma_t}=\pi^{ab}\delta h_{ab}\,d^dx.
\end{equation}
Taking an antisymmetrised variation of the previous expression and integrating over $\Sigma_t$ it is found that
\begin{equation}\label{eq:symplectic_form_GR}
    \Omega(\delta_1,\delta_2)=\int_{\Sigma}d^dx(\delta_1\pi^{ab}\delta_2h_{ab}-\delta_2\pi^{ab}\delta_1h_{ab}).
\end{equation}
We display the bulk canonical expressions here; boundary and corner contributions must be included when required by the chosen boundary conditions \cite{Harlow:2019yfa}.

\section{First law of entanglement entropy}\label{app:B}
In this appendix we will apply the transport prescription to rederive the first law of entanglement entropy \cite{Faulkner:2013ica,Lashkari:2013koa,Taylor:2021jqi}. To this end, we will select a concrete boundary region in the CFT$_2$ and find the variation $\delta_{\mathcal{A}}$ such that $\delta \mathcal{A}/{4G_N}=\Omega_{\textup{bulk}}(\delta,\delta_{\mathcal{A}})$. Once this is known, we will make use of this result to find the variation of the dual variables in the CFT side to compute the boundary dual of the area functional.
First, we recall that the first law of entanglement entropy states that, given a spatial region $\mathcal A$ inside a CFT characterised by a certain state, a perturbation of this state will provoke a change in the von Neumann entropy equal to the change in the expectation value of the modular Hamiltonian.
\begin{equation}
    \delta S_\mathcal A=\delta \braket{K_{\mathcal A}}.
\end{equation}
For arbitrary states, the modular Hamiltonian is highly non-local and, in general, its actual form is not known. However, for the CFT vacuum and a ball-shaped region $\mathcal A$ of radius $R$, the situation greatly simplifies and acquires the form \cite{Lashkari:2013koa,Casini:2011kv}
\begin{equation}
    K_{\mathcal{A}}=2\pi \int_{\mathcal{A}} d^{d-1}x \frac{R^2-r^2}{2R}T^{tt}.
\end{equation}
In the case of a half-infinite region covering half of the CFT spacetime, it reads \cite{Bisognano:1976za, Witten:2018zxz}
\begin{equation}\label{eq: modular_hamiltonian}
    K_{\mathcal A}=2\pi \int_{\mathcal{A}} d^{d-1}x\ x\ T^{tt},
\end{equation}
where $\mathcal A$ is the region with $x\in[0,\infty)$. 

For our discussion, we will restrict to the latter case, taking the AdS$_3$ vacuum as a background, with metric 
\begin{equation}
    ds^2=\frac{\ell^2}{z^2}(-dt^2+dx^2+dz^2).
\end{equation}
Our CFT region $\mathcal A$ will be located at $t=0$ and $x\in[0,\infty)$, so that the HRT surface is a 1-dimensional curve localised at $t=x=0$.

Firstly, it is necessary to determine which is the variation of the metric generated by the area observable, \textit{i.e.} the vector field $\delta_{\mathcal A}$ such that, plugged into the bulk symplectic form \eqref{eq:symplectic_form_GR}, $\Omega_{\textup{bulk}}(\delta,\delta_{\mathcal A})$ is equal to the variation of the area of the HRT surface $\delta \mathcal A/4G_N$. The fact that this observable is located on a codimension-2 surface significantly hinders the application of the Peierls' formalism. For this reason, we will attempt to find this variation based on the result presented in \cite{Kaplan:2022orm}. Let $\Sigma$ be any Cauchy slice containing the HRT surface, with induced metric $h_{ij}$ and extrinsic curvature tensor $K^{ij}$. Here $i,\,j$ run over the coordinates of the surface $\Sigma$, $(x,z)$. The Poisson bracket between the area of the HRT surface and these two phase space variables are\footnote{Notice that, with respect to \cite{Kaplan:2022orm}, a different definition for $\pi^{ij}$ is employed, differing by a factor of $\sqrt{h}$. Despite the modification of the standard Poisson bracket between $h_{kl}$ and $\pi^{ij}$ that in our case reads $\{h_{kl},\pi^{ij}\}=\delta_{(k}^i\delta_{l)}^j\delta^2(x-y)$, the brackets between the area functional and these two variables remains unchanged.}
\begin{equation}
    \delta_{\mathcal A} h_{ij}=\left\{h_{ij},\frac{\mathcal A}{4G_N}\right\}=0,\quad \delta_{\mathcal{A}} K^{ij}=\left\{K^{ij},\frac{\mathcal{A}}{4G_N}\right\}=2\pi \frac{\sqrt{q}}{\sqrt{h}}\delta_\Sigma(\gamma,x) \perp^i\perp^j.
\end{equation}
In the previous expression, $q_{ij}$ is the metric induced on the HRT surface, $\perp$ is the vector normal to the HRT surface within $\Sigma$ and $\delta_\Sigma(\gamma, x)$ is a delta function supported on the HRT surface. Under this convention, $\int_\Sigma d^2 x f(x)\delta_\Sigma (x)=\int_{HRT} dw f(w)$, with $w$ a coordinate on the HRT surface. For the case under study, the Cauchy surface $\Sigma$ will be chosen to be the $t=0$ slice. Considering that the vector normal to the HRT surface within $\Sigma$ is $\perp^i=\frac{z}{\ell}\delta_x^i$, the previous variations reduce to
\begin{equation}\label{eq: variations_area}
    \delta_{\mathcal A} h_{ij}=0,\quad \delta_{\mathcal A} K^{ij}=\frac{2\pi z^3}{\ell^3}\delta(x) \delta^i_x\delta_x^j.
\end{equation}
Once the variation in phase space of these variables is known, one must find a deformation of the AdS$_3$ metric satisfying these two constraints. In particular, this can be achieved using the diffeomorphism generated by the vector field
\begin{equation}
    \xi= 2\pi \Theta(x)(x\partial_t+t\partial_x)
\end{equation}
where $\Theta(x)$ is a Heaviside function supported on the region $x\in[0,\infty)$. Notice that, ignoring this $\Theta(x)$ function, the vector field is the generator of boosts \cite{Headrick:2019eth, Casini:2011kv, Witten:2018zxz, Kaplan:2022orm} and is a Killing vector of AdS$_3$. Thus, if we naively compute the variation it generates on the metric, by construction, it should vanish, so it would be impossible to find \eqref{eq: variations_area}.  This is where the Heaviside function comes into play, since it creates a non-trivial contribution entirely localised at the HRT surface. Indeed, the variation of the metric is zero except for the $xx$ component, which is equal to 
\begin{equation}
    \delta_\xi g_{xx}=\frac{4\pi\, \ell^2\, t\, \delta(x)}{z^2}.
\end{equation}

Plugging it into the  expression for the variations of both $h_{ij}$ and $K^{ij}$ \cite{Herzog:2017xha}

\begin{equation}
    \begin{split}
    \delta_\xi h_{ij}=&\left.\frac{\partial x^\mu}{\partial \sigma^i}\frac{\partial x^\nu}{\partial \sigma^j}\delta_\xi g_{\mu\nu},\right|_\Sigma\\
    \delta_\xi K_{ij}=&\frac{\partial x^\mu}{\partial \sigma^i}\frac{\partial x^\nu}{\partial \sigma^j}\left[-\frac{1}{2}K_{\mu\nu}\delta_\xi g_{nn}-(n_\mu K_\nu^\lambda+n_\nu K_\mu^\lambda)\delta_\xi g_{\lambda n}\right. \\
    &\left.\left.-\frac{1}{2}h_\mu^\lambda h_{\nu}^\rho n^\alpha\left(\nabla_\lambda \delta_\xi g_{\alpha\rho}+\nabla_\rho \delta_\xi g_{\alpha\lambda}-\nabla_\alpha \delta_\xi g_{\lambda\rho}\right)\right]\right|_\Sigma,\\
    \delta_\xi K^{ij}=&\delta_\xi\left.\left(h^{ik}h^{jl}K_{kl}\right)\right|_{\Sigma}
    \end{split}
\end{equation}
yields the expression in \eqref{eq: variations_area}. In the previous formula, the index $n$ refers to a coordinate normal to the surface and $\mu,\nu$ run over all indices. This shows that the diffeomorphism generated by $\xi$ coincides with the flow of the area and then $\delta_\xi \equiv \delta_\mathcal A$.

Once the generator of the area flow is known, it has to be switched to Euclidean signature to read the variation induced on the boundary metric and the vacuum expectation value of the boundary stress tensor. As $t\rightarrow -i\tau$, the vector field becomes
\begin{equation}
    \xi=2\pi i\Theta(x)(x\partial_{\tau}-\tau \partial_x).
\end{equation}
Thence, the variation of the Euclidean metric reads
\begin{equation}
    \delta_\xi g_{xx}=-\frac{4 i \pi\, \ell^2\tau\, \delta (x)}{z^2}
\end{equation}
while the rest of the terms vanish. 

This deformation is still in the Fefferman-Graham gauge and from it, we can read $\delta_{\mathcal A} g^{(0)}_{xx}=- 4i \pi  \tau \delta (x)$  (again, the rest of the components vanish) and $\delta_{\mathcal A} g^{(2)}_{ij}=0$, from what we infer that $\delta_{\mathcal{A}} \gamma_{ij}=\delta_\mathcal{A} g^{(0)}_{ij}$ and $\delta_{\mathcal{A}} \braket{T^{ij}}=0$ \cite{deHaro:2000vlm}.

Substituting into the boundary definition of the symplectic form in \eqref{eq: boundary_symplectic_form} one finds
\begin{equation}\label{eq:boundary simplectic form area}
\begin{split}
    \Omega_{\partial}(\delta,\delta_{\mathcal{A}})&=-\frac{i}{2}\int_{\mathcal{B}_+}d^2 x\sqrt{\gamma}(\delta\braket{T^{xx}}\delta_\mathcal A \gamma_{xx}^+-\delta_\mathcal{A}\braket{T^{xx}}\delta \gamma_{xx}^+)\\
    &=-2\pi \int_{\mathcal{B}_+}d^2 x\ \tau \ \delta(x) \delta\braket{T^{xx}}.
\end{split}
\end{equation}
Let $\zeta=2\pi i\Theta(x)(x\partial_\tau- \tau\partial_x)$ be the projection of $\xi$ to the boundary and $J^{a}=\zeta_b \delta \braket{T^{ab}}$ a current defined in the CFT. Provided that $\delta \braket{T^{ij}}$ is a conserved tensor\footnote{Although $\delta$ may vary the boundary metric, which sources $T^{ij}$, $\delta\braket{T^{ij}}$ is conserved at linear order around the vacuum. Linearising the Ward identity $\nabla_i\braket{T^{ij}} 
\propto\sum_a\braket{\mathcal O^a}\nabla^j\lambda_a$ around vacuum flat space yields an expression whose terms are multiplied by a background quantity. All of them vanish for the vacuum on flat space.}, $\partial_i \delta \braket{T^{ij}}=0$ and the divergence of the former current reads 
\begin{equation}
    \partial_a J^a=\partial_a \zeta_b\delta \braket{T^{ab}}=\frac{1}{2}\delta \braket{T^{ab}}(\partial_a \zeta_b+\partial_b \zeta_a)=\frac{1}{2}\delta \braket{T^{ab}}\mathcal{L}_\zeta \gamma_{ab}=-2\pi i \tau\delta (x)\delta \braket{T^{xx}},
\end{equation}
which, apart from a factor $i$, coincides with the integrand in \eqref{eq:boundary simplectic form area}. In virtue of Stokes' theorem, if $\delta \braket{T^{ab}}$ is assumed to vanish at $\tau\rightarrow \infty$ and $|x|\rightarrow\infty$ the symplectic form can be written as the integral over the $\tau=0$ slice. 
\begin{equation}
    \Omega_\partial(\delta,\delta_{\mathcal A})=-i\left.\int_{-\infty}^\infty dx n_a J^a\right|_{\tau=0}=-2\pi  \int_{-\infty}^\infty dx \ \Theta(x)\ x\ \delta\braket{T^{\tau\tau}}=-2\pi  \int_{0}^\infty dx \ x\ \delta\braket{T^{\tau\tau}},
\end{equation}
with the outward normal of $\mathcal{B}_+$ given by $n_a=-\delta_a^\tau$. As $\braket{T^{\tau\tau}}=-\braket{T^{tt}}$ when continuing $\tau$ to Lorentzian signature, the right-hand side coincides with \eqref{eq: modular_hamiltonian}, so it can be concluded that the boundary symplectic form evaluated for the variation generated by the area is equal to the variation of the expectation value of the modular Hamiltonian for a half-infinite region. If we look at the whole thread of calculations
\begin{equation}
    \frac{\delta \mathcal{A}}{4G_N}=\Omega_{\textup{bulk}}(\delta,\delta_{\mathcal{A}})=\Omega_{\partial}(\delta,\delta_\mathcal{A})=\delta\braket{K_{\mathcal{A}}},
\end{equation}
providing in this way an alternative derivation of $\delta \mathcal{A}/4G_N=\delta \braket{K_{\mathcal A}}$ which, combined with the RT formula, yields the first law of entanglement entropy \cite{Jafferis:2015del}. Although for this computation we have employed the simplest setup, it can be easily extended to other scenarios. For instance, as the calculation has been performed on an AdS$_3$ spacetime, any HRT surface can be related to the one employed here with a suitable isometry. Applying the corresponding transformation to the vector field $\xi$ and repeating the same steps, the same result could be found. For higher dimensions and a half-infinite region, the vector field would read the same and we expect the same computation would go through. However, for a ball-shaped region of radius $R$, we expect that replacing $\xi$ by the generator of boosts within the entangling region multiplied by a step function supported on the portion of the slice contained within the HRT surface and the boundary leads to the same result, but we leave this for future work.

\section{Fefferman-Graham gauge for the York deformation}\label{app:FG_York}

In the Poincar\'e coordinates of Section~\ref{sec3: CV}, the York vector reads \cite{Carrasco:2023fcj}
\begin{equation}
	Y=-\beta_R\left(\frac{z^2}{\sqrt{z^2-t^2}}\,\partial_t+\frac{z\,t}{\sqrt{z^2-t^2}}\,\partial_z\right),
\end{equation}
and, after the continuation $t\to-it_E$,
\begin{equation}
	Y=-i\beta_R\left(\frac{z^2}{\sqrt{z^2+t_E^2}}\,\partial_{t_E}-\frac{z\,t_E}{\sqrt{z^2+t_E^2}}\,\partial_z\right).
\end{equation}
The metric variation it induces is
\begin{equation}
	\begin{split}
		\mathcal{L}_{Y} g_{t_Et_E}&=-\frac{2i\beta_R\, t_E^3}{z^2 (z^2+t_E^2)^{3/2}},\qquad \mathcal{L}_Y g_{zz}=-\frac{2i \beta_R\, t_E}{(z^2+t_E^2)^{3/2}},\\
		\mathcal{L}_{Y} g_{zt_E}&=-\frac{2i\beta_R\, t_E^2}{z\, (z^2+t_E^2)^{3/2}},\qquad \mathcal{L}_Y g_{ab}=-\frac{2i \beta_R\, t_E}{z^2\sqrt{z^2+t_E^2}}\, \delta_{ab},
	\end{split}
\end{equation}
which is not in Fefferman-Graham gauge, since $\mathcal L_Yg_{zz}$ and $\mathcal L_Yg_{zt_E}$ do not vanish. We restore the gauge through the change of coordinates
\begin{equation}\label{eq: change_coords}
	t_E=\tilde t_E+\sum_{n=1}^\infty \tilde z^{2n}f_{2n}(\tilde t_E),\qquad
	z=\tilde z+\sum_{n=1}^\infty \tilde z^{2n+1}g_{2n+1}(\tilde t_E),
\end{equation}
and impose $g_{\tilde z\tilde t_E}=0$ and $g_{\tilde z\tilde z}=1/\tilde z^2$ order by order in $\tilde z$. At the first non-trivial order this gives
\begin{equation}
	f_{2}(\tilde t_E) = \frac{i\beta_R}{\tilde t_E}\,{\rm sign}(\tilde t_E),\qquad
	g_{3}(\tilde t_E) = \frac{i\beta_R}{2\tilde t_E^2}\,{\rm sign}(\tilde t_E).
\end{equation}
In the new coordinates, the leading coefficient of the metric gives $\delta_V\gamma_{ab}=-2i\beta_R\,{\rm sign}(\tilde t_E)\,\gamma_{ab}$. Near the boundary, the combined transformation therefore acts as the rescaling generated by $Y_{\rm bdry}=i\beta_R\,\tilde z\,{\rm sign}(\tilde t_E)\,\partial_{\tilde z}$, as found in~\cite{Carrasco:2023fcj}. Including higher orders in~\eqref{eq: change_coords}, one finds $\delta_V g_{(d)ab}=0$, so that $\delta_V\braket{T_{ab}}=0$ up to the anomalous contribution~\cite{deHaro:2000vlm}.

\bibliographystyle{JHEP}
\bibliography{refs}
\end{document}